\documentclass[aps,prc,twocolumn,floatfix,showpacs,a4paper,nofootinbib]{revtex4}

\usepackage{graphicx} % Include figure files
\usepackage{dcolumn} % Align table columns on decimal point
\usepackage{bm}         % bold math
\usepackage{amsmath}
\usepackage{slashed}
\usepackage{graphicx}
\usepackage{rotating}
\usepackage{float}

\newcommand{\ba}{\begin{eqnarray}}
\newcommand{\ea}{\end{eqnarray}}
\newcommand{\bd}{\begin{displaymath}}
\newcommand{\ed}{\end{displaymath}}
\newcommand{\Lagr}{\mathcal{L}}
\newcommand{\no}{\nonumber}

\newcommand{\sfrac}[2]{{\textstyle\frac{#1}{#2}}}

\def\bra#1{\left\langle #1\right|}
\def\ket#1{\left| #1\right\rangle}

\begin{document}

\title{The $\bar{K} N \rightarrow K \Xi$ reaction in coupled channel chiral models up to next-to-leading order}

\author{
A.~Feijoo, V.K.~Magas and A.~Ramos} 

\affiliation{
Departament d'Estructura i Constituents de la Mat\`eria and Institut de Ci\`encies del Cosmos, Universitat de Barcelona, Mart\'i Franqu\`es 1, E08028 Barcelona, Spain
}

\date{\today}
%\date{19 November 2014}

\begin{abstract}

The meson-baryon interaction in s-wave in the strangeness S=-1 sector has been studied, employing a chiral SU(3) Lagrangian up to next-to-leading order (NLO) and implementing unitarization in coupled channels. The parameters of the Lagrangian have been fitted to a large set of experimental data in different two-body channels, paying special attention to the $\bar{K} N \rightarrow K \Xi$ reaction, which is particularly sensitive to the NLO terms. With the aim of improving the model in the $K\Xi$ production channels, effects of the high spin hyperon resonances $\Sigma(2030)$ and $\Sigma(2250)$ have been taken into account phenomenologically. 

\end{abstract}
\maketitle

\section{Introduction}

Quantum Chromodynamics (QCD) is a gauge theory which describes the strong interaction between the colour charged particles, quarks and gluons. It is strongly coupled at low energies and it cannot be applied perturbatively to describe the interaction of hadrons in this regime. One may then resort to effective theories, such as SU(3) Chiral Perturbation Theory ($\chi$PT), which is built in terms of an effective Lagrangian with hadrons as the new degrees of freedom. This effective theory respects the symmetries of QCD, in particular chiral symmetry $SU(3)_R\times SU(3)_L$ and, more specifically, spontaneous chiral symmetry-breaking that causes the appearance of the Nambu-Goldstone (NG) bosons as light pseudoscalar mesons and the dynamical mass generation of hadrons \cite{Gasser:1983yg,Meissner:1993ah,Bernard:1995dp,Ecker:1994gg,Pich:1995bw}.
While $\chi$PT describes very satisfactorily hadron interactions at low energies, it fails in the vicinity of resonances, which are poles of the scattering amplitude, making the use of nonperturbative schemes mandatory. 

Unitarized Chiral Perturbation Theory (U$\chi$PT), which combines chiral dynamics with unitarization techniques in coupled channels, has shown to be a very powerful tool that permits extending the validity of $\chi$PT to higher energies and to describe the physics around certain resonances, the so called dynamically generated resonances (see \cite{ollerreport} and references therein). A clear example of the success of U$\chi$PT is  the description of the $\Lambda(1405)$ resonance, located only $27$ MeV below the $\bar{K} N$ threshold, that emerges from coupled-channel meson-baryon re-scattering in the $S=-1$ sector. In fact, the dynamical origin of the $\Lambda(1405)$ resonance was already hindered more than 50 years ago \cite{L1405}, an idea that was reformulated later in terms of the chiral unitary theory in coupled channels \cite{KSW}. This success stimulated a lot of activity in the community, which analyzed the effects of including a complete basis of meson-baryon channels, differences in the regularization of the equations, s- and u-channel Born terms in the Lagrangian, next-to-leading (NLO) contributions, etc \dots \cite{KWW,OR,OM,LK,BMW,2pole,BFMS,BNW,BMN}. The various developed models could reproduce the $\bar{K} N$ scattering data very satisfactorily and all these efforts culminated in establishing the $\Lambda(1405)$ as a superposition of two poles of the scattering amplitude  \cite{OM,2pole,PRL}, generated dynamically from the unitarized meson-baryon interaction in coupled channels.

This topic experienced a renewed interest in the last few years, after the availability of a more precise measurement of the energy shift and width of the 1s state in kaonic hydrogen by the SIDDHARTA collaboration \cite{SIDD} at DA$\Phi$NE. 
The CLAS collaboration at JLAB has also recently provided mass distributions of $\Sigma^+ \pi^-$, $\Sigma^- \pi^+$, and $\Sigma^0 \pi^0$ states in the region of the $\Lambda(1405)$ \cite{Moriya:2013eb}, as well as differential cross sections \cite{Moriya:2013hwg} and a direct determination of the expected spin-parity $J^\pi=1/2^-$ of the $\Lambda(1405)$\cite{Moriya:2014kpv}. Invariant $\pi\Sigma$ mass distributions from $pp$ scattering experiments have recently been  measured by the COSY collaboration at J\"ulich \cite{Zychor:2007gf} and by the HADES collaboration at GSI \cite{Agakishiev:2012xk}. In parallel with the increased experimental activity, the theoretical models have been revisited \cite{IHW,HJ_rev,GO,MM,crimea,MFST} and analyses of the new reactions, aiming at pinning  down the properties of the $\Lambda(1405)$ better, have been performed \cite{Roca:2013av,Roca:2013cca,Mai:2014xna}. 

In this paper, we present a study of the $S=-1$ meson-baryon interaction which aims at providing well constrained values of the low-energy constants of the NLO chiral Lagrangian. We will employ data in the strong sector, including elastic and inelastic cross section data ($K^- p \to K^- p$, ${\bar K}^0 n$, $\pi^{\pm}\Sigma^{\mp}$, $\pi^0\Sigma^0$, $\pi^0\Lambda$) and the precise SIDDHARTA value of the energy shift and width of kaonic hidrogen,  as done by the recent works, but, in addition, we will also constrain the parameters of our model to reproduce the  $K\Xi$ production data via the reactions $K^- p\to K^+\Xi^-, K^0\Xi^0$. The motivation lies in the fact that, as we will see, 
the lowest-order Lagrangian does not contribute directly to these reactions, which then become especially sensitive to the NLO terms. In the first part of this paper, we will present results that support the idea that the $K\Xi$ cross sections are crucial ingredients for determining the values of the low-energy constants of the NLO Lagrangian.

We will also show that certain structures present in the $K^- p\to K^+\Xi^-, K^0\Xi^0$ cross sections cannot be accounted for by the unitary coupled-channels model at NLO. The contribution of explicit resonant terms is an unavoidable fact at CM energies of around 2 GeV characteristic of $K\Xi$ production. In fact, several  resonance-based models have investigated the photoproduction of $\Xi$ particles off the proton \cite{Nakayama:2006ty,Man:2011np} or via the strong reactions $K^- p\to K^+\Xi^-, K^0\Xi^0$ investigated here \cite{Sharov:2011xq,Shyam:2011ys,Jackson:2014hba,jackson}. These works have found a non-negligible contribution from high-spin hyperon resonances. Guided by these findings, in a second part of this paper we implement phenomenological resonant contributions from high spin resonances, specifically  $\Sigma(2030)$ and $\Sigma(2250)$, which were those found to contribute more strongly in the study of Ref.~\cite{Sharov:2011xq}. This is a necessary exercise to establish how much strength in the $K^- p\to K^+\Xi^-, K^0\Xi^0$ reactions can already be described by resonant terms and how much background, accounted here by the terms of the NLO Lagrangian, is then needed for a  good reproduction of data. We will see that, even with resonances explicitly included, the NLO Lagrangian is an important and necessary ingredient of the model.

This work is organized in two parts. Section \ref{chiral} is devoted to the contributions of the chiral Lagrangian. After presenting briefly the formalism of unitarization in coupled channels, the procedure adopted for determining the parameters of the model is described, followed by a discussion of the results obtained with the various orders of the chiral Lagrangian.  In a similar way, Section~\ref{reson} starts with a description of the formal technicalities involved in the inclusion of the resonances, after which their ability in reproducing the $\Xi$ strong production data and their effect on the background terms implemented by the chiral Lagrangian are discussed. A summary and some concluding remarks are given in the Section \ref{conclusions}.

\section{Chiral unitary approach}
\label{chiral}

\subsection{Formalism} 

In this section we present in detail the coupled-channel formalism employed for describing  meson-baryon scattering. The starting point is the SU(3) chiral effective Lagrangian, 
\begin{equation}
 \label{Lagr}
\Lagr = \Lagr_\phi + \Lagr_{\phi B} \ ,    
\end{equation} 
which incorporates the same symmetries and chiral spontaneous symmetry breaking patterns as QCD, and describes the coupling of the pseudoscalar octet $(\pi,K,\eta)$ to the fundamental baryon octet $(N,\Lambda,\Sigma,\Xi)$.
The first term takes into account pure mesonic processes, while the $\Lagr_{\phi B}$ term, of interest for this work, implements the interactions between mesons and baryons. At lowest order, it reads
\ba \label{LagrphiB1}
\Lagr_{\phi B}^{(1)} & = & i \langle \bar{B} \gamma_{\mu} [D^{\mu},B] \rangle
                            - M_0 \langle \bar{B}B \rangle 
                            - \frac{1}{2} D \langle \bar{B} \gamma_{\mu}
                             \gamma_5 \{u^{\mu},B\} \rangle \no \\
                     &   &  - \frac{1}{2} F \langle \bar{B} \gamma_{\mu} 
                               \gamma_5 [u^{\mu},B] \rangle \ ,
\ea
where $M_0$ is the common baryon octet mass in the chiral limit, the constants $D$, $F$ denote the axial vector couplings of the baryons to the mesons, and the symbol $\langle \dots \rangle$ stands for the trace in flavor space.

The pseudoscalar meson octet $\phi$ is arranged in a matrix valued field
\begin{equation}
\label{Uphi}
U(\phi) = u^2(\phi) = \exp{\left( \sqrt{2} i \frac{\phi}{f} \right)},
\end{equation} 
with
\begin{equation}
\label{mesfield}
\phi=\begin{pmatrix}
\frac{1}{\sqrt{2}}\pi^0+\frac{1}{\sqrt{6}}\eta & \pi^+ & K^+\\
\pi^- & -\frac{1}{\sqrt{2}}\pi^0+\frac{1}{\sqrt{6}}\eta & K^0\\
 K^- & \bar{K}^0 & -\frac{2}{\sqrt{6}}\eta\\
\end{pmatrix},
\end{equation} 
and $f$ being the pseudoscalar decay constant in the chiral limit. The quantity $U$ enters the  Lagrangian in the combinations $u_\mu = i u^\dagger \partial_\mu U u^\dagger$.

The octet baryon fields are collected in
\begin{equation}
\label{Bfield}
B=\begin{pmatrix}
\frac{1}{\sqrt{2}}\Sigma^0+\frac{1}{\sqrt{6}}\Lambda & \Sigma^+ & p\\
\Sigma^- & -\frac{1}{\sqrt{2}}\Sigma^0+\frac{1}{\sqrt{6}}\Lambda & n\\
\Xi^- & \Xi^0 & -\frac{2}{\sqrt{6}}\Lambda\\
\end{pmatrix}  \ ,
\end{equation} 
and, finally, $[D_\mu, B]$ stands for the covariant derivative
\begin{equation}
\label{CoDer}
[D_\mu, B] = \partial_\mu B + [ \Gamma_\mu, B] \ ,
\end{equation} 
where the chiral connection $\Gamma_\mu$ is given by
\begin{equation}
\label{Conn}
\Gamma_\mu = \sfrac{1}{2} [ u^\dagger,  \partial_\mu u] . 
\end{equation} 
From the Lagrangian $\Lagr_{\phi B}^{(1)}$, one can derive the meson-baryon interation kernel at lowest order, or Weinberg-Tomozawa (WT) term, which  reads:
\begin{equation}
\label{WT}
V^{\scriptscriptstyle WT}_{ij}=-C_{ij}\frac{1}{4f^2}\bar{u}_{B_j}^{s'}(p_j)\gamma^\mu u_{B_i}^{s}(p_i)(k_{i\mu}+k_{j\mu})\ ,
\end{equation} 
and depends only on one parameter, the pion decay constant $f$. The indices $(i,j)$ cover all the initial and final channels, which, in the case of strangeness $S=-1$ and charge $Q=0$ explored here, amount to ten: $K^-p$, $\bar{K}^0 n$, $\pi^0\Lambda$, $\pi^0\Sigma^0$, $\pi^-\Sigma^+$, $\pi^+\Sigma^-$, $\eta\Lambda$, $\eta\Sigma^0$, $K^+\Xi^-$ and $K^0\Xi^0$. The matrix of coefficients $C_{ij}$ is shown in Appendix \ref{appendix}, see Table \ref{tab7}.  The four-momenta $k_{i\mu}$ and  $k_{j\mu}$ are those of the incoming and outgoing mesons, respectively, while $u_{B_i}^{s}(p_i)$ denotes the incoming baryon spinor with spin $s$  and momentum $p_i$, and analogously for the spinor $\bar{u}_{B_j}^{s'}(p_j)$ of the outgoing baryon. The pion decay constant is well known experimentally, $f_{\rm exp}=93.4$~MeV, however in LO U$\chi$PT calculations this parameter is usually taken to be $f=1.15-1.2 f_{\rm exp}$, in order to partly simulate the effect of the higher order corrections.  We will leave it as a parameter of our fits.

At next-to-leading order, the contributions of $\Lagr_{\phi B}$ to meson-baryon scattering are:
\begin{eqnarray}
    \Lagr_{\phi B}^{(2)}& = & b_D \langle \bar{B} \{\chi_+,B\} \rangle
                             + b_F \langle \bar{B} [\chi_+,B] \rangle
                             + b_0 \langle \bar{B} B \rangle \langle \chi_+ \rangle \no \\ 
                        &   &+ d_1 \langle \bar{B} \{u_{\mu},[u^{\mu},B]\} \rangle
                             + d_2 \langle \bar{B} [u_{\mu},[u^{\mu},B]] \rangle    \no \\ 
                        &   &+ d_3 \langle \bar{B} u_{\mu} \rangle \langle u^{\mu} B \rangle
                             + d_4 \langle \bar{B} B \rangle \langle u^{\mu} u_{\mu} \rangle \ ,
\label{LagrphiB2}
\end{eqnarray}
where $\chi_+ = 2 B_0 (u^\dagger \mathcal{M} u^\dagger + u \mathcal{M} u)$ breaks chiral symmetry explicitly via the quark mass matrix  $\mathcal{M} = {\rm diag}(m_u, m_d, m_s)$ and $B_0 = - \bra{0} \bar{q} q \ket{0} / f^2$ relates to the order parameter of spontaneously broken chiral symmetry.
The coefficients $b_D$, $b_F$, $b_0$ and $d_i$ $(i=1,\dots,4)$ are the low energy constants. In principle, the first three coefficients, involved in terms proportional to the $\chi_+$ field,  should fulfill constraints related to the mass splitting of baryons. However, since our study  goes beyond tree level and incorporates higher order terms via coupled channel scattering equations, we will relax those constraints and will fit these $b$-type constants, together with the $d_i$ ones, to the experimental data. 

From the Lagrangian $\Lagr_{\phi B}^{(2)}$ one can derive the meson-baryon interaction kernel at NLO:
\begin{equation}
\label{V_NLO}
V^{\scriptscriptstyle NLO}_{ij}=\frac{1}{f^2}\bar{u}_{B_j}^{s'}(p_j)\left(D_{ij}-2(k_{i\mu}k^{j\mu})L_{ij}\right)u_{B_i}^{s}(p_i),
\end{equation} 

The interaction kernel up to NLO is taken in the non-relativistic limit and reads
$$
V_{ij}=V^{\scriptscriptstyle WT}_{ij}+V^{\scriptscriptstyle NLO}_{ij}=
$$
\begin{equation}
\label{V_TOT}
 - \frac{C_{i j}(2\sqrt{s} - M_{i}-M_{j})}{4 f^2}\! N_{i} N_{j}
+
\frac{D_{ij}-2(k_\mu k^{\prime\,\mu})L_{ij}}{f^2}\! N_{i} N_{j}\,.
\end{equation}

where
\begin{equation}
N_{i}=\sqrt{\frac{M_i+E_i}{2M_i}},\,\, N_{j}=\sqrt{\frac{M_j+E_j}{2M_j}} \no ,
\end{equation}
and $M_i,M_j$ and $E_i,E_j$ are the masses and energies, respectively, of the baryons involved in the transition. The
$D_{ij}$ and $L_{ij}$ coefficients depend on the new parameters $b_0$, $b_D$, $b_F$, $d_1$, $d_2$, $d_3$ and $d_4$ and are given in Appendix \ref{appendix}, see Table \ref{tab8}.

\begin{figure}[ht]
 \centering
 \includegraphics[width=3.5in]{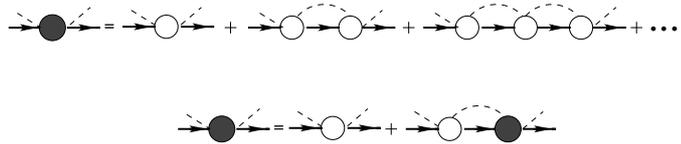}
%\vskip -25mm
 \caption{ 
Diagrammatic solution of the Bethe-Salpeter equation, where the interaction kernel $V$ is represented by the empty blobs, the scattering matrix $T$ - by the solid blobs, and the loop function  $G$ is represented by the intermediate baryon-meson propagators.
}
%\vskip -2mm
 \label{fig1}
\end{figure}

As mentioned in the Introduction,  the interaction kernel cannot be employed perturbatively to describe
the scattering of $\bar{K}N$ states, since they couple strongly to many other channel states and generate the $\Lambda(1405)$ resonance close to their threshold. Due to this reason, a nonperturbative resummation is needed. The U$\chi$PT method consists in solving the Bethe-Salpether equation in coupled channels, using the interaction kernel derived from the chiral Lagrangian. The corresponding equation for the scattering amplitude $T_{ij}$ is schematically displayed in Fig.~\ref{fig1}
and corresponds to the infinite sum 
\begin{equation}
\label{LS_expanded}
T_{ij} =V_{ij}+V_{il} G_l V_{lj}+V_{il} G_l V_{lk}G_k V_{kj}+... \  ,
\end{equation} 
where the subscripts $i,j,l,\dots$ run over all possible channels and the loop function $G_i$ stands for the propagator of the meson-baryon state of channel $i$. The former equation can also be cast as:
\begin{equation}
T_{ij} =V_{ij}+V_{il} G_l T_{lj}  \ .
\label{LS}
\end{equation} 
This is an integral equation for the amplitude $T_{ij}$ which, in the CM frame, involves an integral over the four-momentum of the intermediate  meson-baryon system, which can take off-shell values. However, it has been shown \cite{OR,HJ_rev,OO_ND} that the off-shell parts of the interaction kernel gives rise to tadpole-type diagrams, which can be reabsorbed into renormalization of couplings and masses and could hence be omitted from the calculation. This procedure permits factorizing $V_{il}$ and $T_{lj}$ out of the integral equation, leaving a simple system of algebraic equations to be solved, which, in matrix form reads:
\begin{equation}
T ={(1-VG)}^{-1}V ,
 \label{T_algebraic}
\end{equation} 
%More details of the application of the method, for instance, are shown Ref. %\cite{OO,OR,HJ_rev}. 
%For completeness, we should mention that there is another, less intuitive, method %to derive the amplitude as the one obtained by LS equation; it is the N/D method %based in the optical theorem \cite{OO_ND, MO_ND, OM}. Hence, the amplitude %can be found by solving the algebraic matrix equation
where the loop function $G$ stands for a diagonal matrix with elements: 
\begin{equation} \label{Loop_integral}
G_l={\rm i}\int \frac{d^4q_l}{{(2\pi)}^4}\frac{2M_l}{{(P-q_l)}^2-M_l^2+{\rm i}\epsilon}\frac{1}{q_l^2-m_l^2+{\rm i}\epsilon} \ ,
\end{equation} 
where $M_l$ and $m_l$ are the baryon and meson masses of the $``l"$ channel. The loop function diverges logarithmically and needs to be regularized. We apply dimensional regularization, which gives:
\ba
& G_l = &\frac{2M_l}{(4\pi)^2} \Bigg \lbrace a_l+\ln\frac{M_l^2}{\mu^2}+\frac{m_l^2-M_l^2+s}{2s}\ln\frac{m_l^2}{M_l^2} + \no \\ 
 &     &\frac{q_{\rm cm}}{\sqrt{s}}\ln\left[\frac{(s+2\sqrt{s}q_{\rm cm})^2-(M_l^2-m_l^2)^2}{(s-2\sqrt{s}q_{\rm cm})^2-(M_l^2-m_l^2)^2}\right]\Bigg \rbrace ,  
 \label{dim_reg}    
\ea 
where $\mu$ is the dimensional regularization scale (we take $\mu=1$ GeV), and $a_l$ are the so called subtraction constants. They will be used as free parameters and fitted to the experimental data. Taking into account isospin symmetry, there are only 6 independent subtraction constants in the $S=-1$ meson-baryon scattering problem studied here.

We note that, at lowest order in the chiral expansion, there is also the contribution of the $s$ and $u$-channel diagrams involving the coupling of the meson-baryon channel to an intermediate baryon state. The contribution of these terms is very moderate \cite{OM} and, although they have been shown to help in producing more physical values of the subtracting constants in some cases \cite{BMN,IHW}, they do not influence significantly the quality of the fits. We have neglected these terms in the present study. 

Once the $T$-matrix is known, we can obtain the unpolarized differential and total cross-section for the $i \rightarrow j$ reaction:
\begin{equation}
\sigma_{ij}=\frac{1}{4\pi}\frac{M_iM_j}{s}\frac{k_j}{k_i}S_{ij} \ ,
 \label{sigma}
\end{equation}
where  $s$ is the square of the center-of-mass (CM) energy, and we have averaged over the initial baryon spin projections and resummed over the final baryon spin projections:
\begin{equation}
\quad S_{ij}=\frac{1}{2}\sum_{s',s}|T_{ij}(s',s)|^2 \ .
 \label{M_matrix}
\end{equation}
The $K^- p$ scattering length is obtained from the $K^- p$ scattering amplitude at threshold as:
\begin{equation}
a_{\scriptscriptstyle K^- p}=-\frac{1}{4\pi}\frac{M_p}{\sqrt{M_p+M_{\bar K}}}T_{K^- p \to K^- p}\ ,
 \label{scat_lenght}
\end{equation}
where we have used the following notation 
\begin{equation}
\quad T_{ij}=\frac{1}{2}\sum_{s',s} T_{ij}(s',s) \ .
 \label{T_matrix}
\end{equation}
The scattering length is related to the energy shift and width of the 1s state of kaonic hydrogen via the second order corrected Deser-type formula \cite{Meissner:2004jr} :
\begin{equation}
\Delta E-i\frac{\Gamma}{2}=-2\alpha^3\mu_{r}^{2} a_{\scriptscriptstyle K^- p} \Big[ 1+2 a_{\scriptscriptstyle K^- p}\,\alpha\,\mu_r\, (1-\ln\alpha) \Big],                                                                                                                       
\label{ener_shift_width}
 \end{equation} 
where $\alpha$ is the fine-structure constant and $\mu_r$ the reduced mass of the $K^-p$ system.

From the elastic and inelastic $K^-p$ cross sections evaluated at threshold, one can also obtain the following measured branching ratios of cross section yields: 
\ba 
\gamma & = & \frac{\Gamma(K^- p \rightarrow \pi^+ \Sigma^-)}{\Gamma(K^- p \rightarrow \pi^- \Sigma^+)}  ,\\ 
 R_n & = & \frac{\Gamma(K^- p \rightarrow \pi^0 \Lambda)}{\Gamma(K^- p \rightarrow {\rm neutral\, states})}   , \\
 R_c & = & \frac{\Gamma(K^- p \rightarrow \pi^+ \Sigma^-,\pi^- \Sigma^+ )}{\Gamma(K^- p \rightarrow {\rm inelastic\, channels})} \ .     
\label{branch_ratios} 
\ea  

\subsection{Data treatment and fits}

We consider a large amount of cross section data for $K^-p$ scattering and related channels \cite{exp_1,exp_2, exp_3, exp_4, exp_5, exp_6, exp_7, exp_8, exp_9, exp_10, exp_11}. Some points of these data sets have inconsistencies with the trend of the neighbouring points and have not been employed in the fitting procedure, leaving us with a total of 161 experimental points coming from $K^-p$ scattering. The points eliminated will be displayed in red in the figures.
We also fit the parameters of our model to the measured branching ratios \citep{br_1,br_2} 
\ba 
\gamma & = & 2.36 \pm 0.04 \no\ ,\\ 
 R_n & = & 0.664 \pm 0.011  \no\ , \\
 R_c & = & 0.189 \pm 0.015 \no\ ,       
\ea 
and to the recent energy shift and width of the 1s state of kaonic hydrogen obtained by the SIDDHARTA Collaboration \cite{SIDD}, namely $\Delta E_{1s}=283\pm36$ and $\Gamma_{1s}=541\pm92$.
The distribution of points per observable is summarized in Table~\ref{tab1}.
\begin{table}[ht]
\begin{tabular}{lc}
\hline \\[-2.5mm]
 {Observable}\, & \, {Points} \\
\hline \\[-2.5mm]
$\sigma_{K^-p \to K^-p}$ & 23 \\
$\sigma_{K^-p \to \bar{K}^0n}$ & 9 \\
$\sigma_{K^-p \to \pi^0\Lambda}$ & 3 \\
$\sigma_{K^-p \to \pi^0\Sigma^0}$ & 3 \\
$\sigma_{K^-p \to \pi^-\Sigma^+}$ & 20 \\
$\sigma_{K^-p \to \pi^+\Sigma^-}$ & 28 \\
$\sigma_{K^-p \to K^+\Xi^-}$ & 46 \\
$\sigma_{K^-p \to K^0\Xi^0}$ & 29 \\
$\gamma$ & 1 \\
$R_n$ & 1 \\
$R_c$ & 1 \\
$\Delta E_{1s}$ & 1 \\
$\Gamma_{1s}$ & 1 \\
\hline
\end{tabular}
\caption{Number of experimental points used in our fits, distributed per observable.}
\label{tab1}
\end{table}

The standard fitting procedure consists of minimizing $\chi^2_{\rm d.o.f.}$, defined as:
\begin{equation}
\label{Chi^2_clas}
\chi^{2}_{\rm d.o.f.}=\frac{1}{\sum_{k=1}^K n_k - p} \sum_{k=1,i=1}^{K,n_k}\frac{\big( y_{i,k}^{\rm th}- y_{i,k}^{\rm exp}\big)^2}{\sigma_{i,k}^{2}}
\end{equation}
where $y_{i,k}^{\rm exp}$, $y_{i,k}^{\rm th}$ and $\sigma_{i,k}$ represent, respectively, the experimental value, theoretical prediction and error of the $i^{th}$ point of the $k^{th}$ observable, which has a total of $n_{k}$ points, $K$ is the total number of observables, and $p$ denotes the number of free fitting parameters. This previous definition could suppress the relevance in the fit of observables which have a small number of associated experimental points, in favour of those with a larger set. To circumvent this problem, we adopt the method already exploited in  \cite{GO, IHW}, which uses a normalized $\chi^2$ that assigns equal weight to the different measurements. This is achieved by averaging over the different experiments the corresponding $\chi^2$ per degree of freedom,  which is obtained by dividing the contribution of the experiment, $\chi^2_k$, by its own number of experimental points, $n_k$. More explicitly, the redefined $\chi^2$ per degree of freedom, which we will use in this work, is given by the expression
\begin{equation}
\chi^{2}_{\rm d.o.f}=\frac{\sum_{k=1}^K n_k }{\left( \sum_{k=1}^K n_k -p \right)} \frac{1}{K} \sum_{k=1}^K \frac{\chi^{2}_{k}}{n_k}
\label{Chi^2_dof}
\end{equation}
with
\begin{equation}
 \chi^{2}_{k}=\sum_{i=1}^{n_k}\frac{\left( y_{i,k}^{\rm th}- y_{i,k}^{\rm exp}\right)^2}{\sigma_{i,k}^{2}} \ .\nonumber 
\end{equation}

\begin{table*}[t]
\begin{tabular}{lcccccc}
\hline \\[-2.5mm]
   & {$\gamma$} & {$R_n$} & {$R_c$} & {$a_p(K^-p \rightarrow K^- p)$} & {$\Delta E_{1s}$} & {$\Gamma_{1s}$} \\
\hline \\[-2.5mm]
WT  (no $K\Xi$)& 2.37 & 0.191 & 0.665 & $-0.76+{\rm i\,}0.79$ & 316 & 511 \\%
NLO (no $K\Xi$)& 2.36 & 0.188 & 0.662 & $-0.67+{\rm i\,}0.84$ & 290 & 559 \\%
WT             & 2.36 & 0.192 & 0.667 & $-0.76+{\rm i\,}0.84$ & 318 & 543\\%
NLO            & 2.36 & 0.189 & 0.664 & $-0.73+{\rm i\,}0.85$ & 310 & 557\\%
\hline \\[-2.5mm]
Exp.           & $2.36$ &	$0.189$ & $0.664$ & $ -0.66 + {\rm i\,}0.81$ &	$283$ & $541$ \\
               & $\pm 0.04$ & $\pm 0.015$ & $\pm 0.011$ & $(\pm0.07)+ {\rm i\,}(\pm0.15)$ &	$\pm36$ & $\pm92$ \\
\hline
\end{tabular}
\caption{Threshold observables obtained from the WT (no $K\Xi$), NLO (no $K\Xi$), WT and NLO fits explained in the text. Experimental data is taken from \cite{SIDD,br_1,br_2}. }
\label{tab:thresh}
\end{table*}
In order to compare with previous works and learn about the importance of the different terms of the chiral Lagrangian, we perform, in this part of the work, four different fits.  The first fit corresponds to a unitarized calculation employing the chiral Lagrangian up to the lowest order WT term. This involves the fitting of seven parameters: the pion decay constant $f$ and the six subtraction constants $a_{\bar{K}N}$,  $a_{\pi\Lambda}$, $a_{\pi\Sigma}$, $a_{\eta\Lambda}$, $a_{\eta\Sigma}$, and $a_{K\Xi}$. The second fit improves upon the first one by using up to the NLO terms of the interaction kernel, thus involving seven additional parameters: the NLO low energy constants $b_0$, $b_D$, $b_F$, $d_1$, $d_2$, $d_3$ and $d_4$.
Analogously to previous works, these first and second fits ignore the  experimental data corresponding to the $K\Xi$ channels and their results will be denoted by WT (no $K\Xi$) and NLO (no $K\Xi$), respectively. From the prediction of the $K\Xi$ cross sections given by these fits, we will demonstrate clearly the important role that the NLO terms have on the $K^- p \to K\Xi$ reactions. This brings up, naturally, the study of the third and fourth fits, denoted by WT and NLO, respectively, which correspond to the same fitting procedures than the first and second ones but including the $K\Xi$ production cross section data. 

\subsection{Results and discussion}
\label{subsec:results1}

In this section, we present the results obtained with the above mentioned fits, namely using the WT kernel of Eq.~(\ref{WT}) or the NLO one of Eq.~(\ref{V_TOT}), and considering or not the experimental data of the $K\Xi$ channels.

We start by showing the results obtained for the threshold observables, collected in Table~\ref{tab:thresh}. It is clear that all the fits are able to reproduce, with a similar degree of accuracy, the branching ratios, the $K^-p$ scattering length and the related energy shift and width of the 1s state of kaonic hydrogen, which is also shown in the table for completeness. Actually, only the first and third fits, obtained with the lowest order WT kernel, seem produce a worse value of the real part of $a_{K^-p}$, close to the limit of its error band. 

\begin{figure*}[!htb]
\centering
 \includegraphics[width=5in]{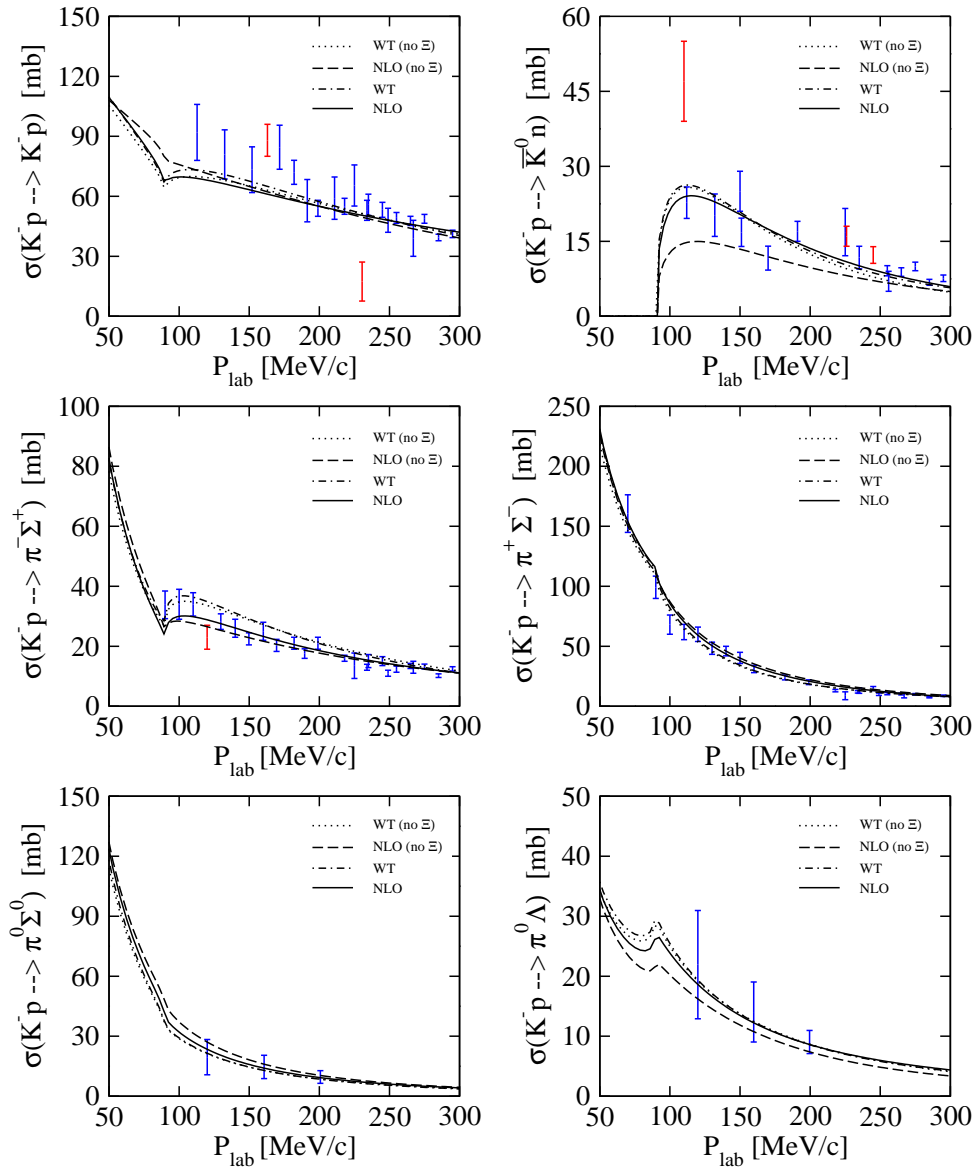}
\caption{Total cross sections for the $K^- p\to K^- p, \bar{K}^0n, \pi^- \Sigma^+, \pi^+\Sigma^-, \pi^0 \Sigma^0, \pi^0\Lambda $ reactions obtained from the WT (no $K\Xi$) fit (dotted line), the NLO (no $K\Xi$) fit (dashed line),  the WT fit (dot-dashed line) and the NLO fit (solid line), where the last two cases take into account the experimental data of the $K\Xi$ channels, see text for more details. Experimental data are from \cite{exp_1,exp_2, exp_3, exp_4}. The points in red have not been included in the fitting procedure.} 
  \label{fig2}
\end{figure*}
\begin{figure}[!htb]
\centering
  \includegraphics[width=2.8in]{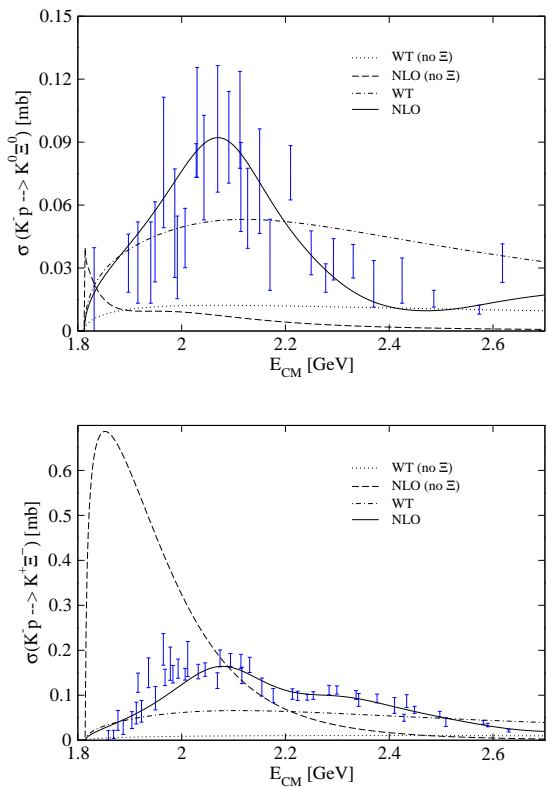}
%\vspace{0.5cm}
\caption{The total cross sections of the $K^- p\to K^0 \Xi^0, K^+ \Xi^-$ reactions obtained from the WT (no $K\Xi$) fit (dotted line), the NLO (no $K\Xi$) fit (dashed line),  the WT fit (dot-dashed line) and the NLO fit (solid line). Experimental data are from \cite{exp_5, exp_6, exp_7, exp_8, exp_9, exp_10, exp_11}.} 
  \label{fig3}
\end{figure}

Similarly, the four fitting schemes reproduce satisfactorily the total cross sections for the reactions $K^- p\to K^- p$, $\bar{K}^0n$, $\pi^- \Sigma^+$, $\pi^+\Sigma^-$, $\pi^0 \Sigma^0$, $\pi^0\Lambda$ shown in Fig.~\ref{fig2}. However, substantial differences appear in the description of the $K^0\Xi^0$, $K^+\Xi^-$ production channels, displayed  in Fig.~\ref{fig3}. The results of the WT (no $K\Xi$) fit, represented by dotted lines, cannot even reproduce the size of the cross section in either reaction. The predicted cross sections amount to less than 0.015 mb, i.e. one order of magnitude smaller than the measured ones. This is not a surprising result, because there is no direct contribution from the reactions $K^- p\to K^0 \Xi^0, K^+ \Xi^-$ at lowest order, since the coefficient $C_{ij}$ in Eq.~(\ref{WT}) is 0 in both cases (see  Table~\ref{tab7} in Appendix \ref{appendix}).
Consequently, the only contribution to the scattering amplitude of these channels comes from the effect of the rescattering terms generated by the coupled channels unitarization, which is not sufficient to reproduce the strength of these cross sections. This fact leads us to believe that these reactions are very sensitive to the NLO corrections, due to non-zero value of the  $L_{K^-p\to K \Xi}$ coefficients of the potential of Eq.~(\ref{V_NLO}) (see table \ref{tab8} in Appendix \ref{appendix}). This is confirmed already by the NLO (no $K\Xi$) results  represented by dashed lines in Fig.~\ref{fig3}. Even if the experimental data for the  $K^- p\to K^0 \Xi^0, K^+ \Xi^-$ reactions have not been employed in this fit, the NLO (no $K\Xi$)  result gives a larger amount of strength for this channels, especially in the case of the $K^+ \Xi^-$  production reaction, where the prediction even overshoots the data considerably. 

The obvious next step is to include the $K\Xi$ data in the fitting procedure and, naturally, the NLO results, represented by the solid line, reproduce quite satisfactorily the  $K^- p\to K^0 \Xi^0, K^+ \Xi^-$ cross sections. For completeness, we have also attempted to reproduce these reactions employing only the lowest order Lagrangian. The corresponding WT results, represented by the dot-dashed lines, improve considerably over those of the WT (no $K\Xi$) fit, but the fact that the lowest order Lagrangian can only affect these channels through unitarization, gives rise to quite unphysical values for the fitted subtraction constants, as commented below.
\begin{table*}[ht]
\begin{tabular}{lrrrr}
\hline \\[-2.5mm]
 &  WT (no $K\Xi$) & NLO (no $K\Xi$)
  &  WT &  NLO \\
\hline \\[-2.5mm]
$a_{\bar{K}N} \ (10^{-3})$&$-1.681\pm 0.738$  &$5.151\pm 0.736$  & $-1.986\pm 2.153$  & $6.550\pm 0.625$  \\
$a_{\pi\Lambda}\ (10^{-3})$    &$ 33.63\pm 11.11$&$21.61\pm 10.00$& $-248.6\pm 122.0$&       $54.84\pm 7.51$ \\
$a_{\pi\Sigma}\ (10^{-3})$     &$0.048\pm 1.925$   &$3.078\pm 2.101$  & $0.382\pm 2.711$   & $-2.291\pm 1.894$ \\
$a_{\eta\Lambda}\ (10^{-3})$   &$1.589\pm 1.160$   &$-10.460\pm 0.432$& $1.696\pm 2.451$   & $-14.16\pm 12.69$ \\
$a_{\eta\Sigma}\ (10^{-3})$    &$-45.87\pm 14.06$&$-8.577\pm 0.353$ & $277.8\pm 139.1$ & $-5.166\pm 0.068$ \\
$a_{K\Xi}\ (10^{-3})$          &$-78.49\pm 47.92$&$4.10\pm 12.67$ & $30.85\pm 10.58$ & $27.03\pm 7.83$ \\
$f/f_{\pi}$                    &$1.202\pm 0.053$   &$1.186\pm 0.012$  & $1.202\pm 0.119$   & $1.197\pm 0.008$  \\
$b_0 \ (GeV^{-1}) $            &   \multicolumn{1}{c}{-}                 &$ -0.861\pm 0.014$& \multicolumn{1}{c}{-}                    & $-1.214\pm 0.014$ \\
$b_D \ (GeV^{-1}) $            &    \multicolumn{1}{c}{-}                &$0.202\pm 0.011$  & \multicolumn{1}{c}{-}                    & $0.052\pm 0.040$  \\
$b_F \ (GeV^{-1}) $            &    \multicolumn{1}{c}{-}                &$0.020\pm 0.057$  & \multicolumn{1}{c}{-}                    & $0.264\pm 0.146$  \\
$d_1 \ (GeV^{-1}) $            &    \multicolumn{1}{c}{-}                &$0.089\pm 0.096$  & \multicolumn{1}{c}{-}                    & $-0.105\pm 0.056$ \\
$d_2 \ (GeV^{-1}) $            &    \multicolumn{1}{c}{-}                &$0.598\pm 0.062$  & \multicolumn{1}{c}{-}                    & $0.647\pm 0.019$  \\
$d_3 \ (GeV^{-1}) $            &    \multicolumn{1}{c}{-}                &$0.473\pm 0.026$  &  \multicolumn{1}{c}{-}                   & $2.847\pm 0.042$  \\
$d_4 \ (GeV^{-1}) $            &    \multicolumn{1}{c}{-}                &$-0.913\pm 0.031$ &  \multicolumn{1}{c}{-}                   & $-2.096\pm 0.024$ \\
\hline \\[-2.5mm]
$\chi^2_{\rm d.o.f.}$                & $0.62$               & $0.39$              & $2.57$                &  $0.65$    \\
\hline
\end{tabular}
\caption{Values of the parameters and the corresponding  $\chi^2_{\rm d.o.f.}$, defined as in Eq.~(\ref{Chi^2_dof}), for the different fits described in the text. The value of the pion decay constant is $f_{\pi}=93$ MeV and the subtraction constants are taken at a regularization scale $\mu=1$ GeV.} 
\label{tab2}
\end{table*}

Table~\ref{tab2} displays the values of the parameters of the four fits discussed in this section, together with the obtained  value of $\chi^2_{\rm d.o.f.}$. 
Note first that the larger value of  $\chi^2_{\rm d.o.f.}$ in the  NLO fit with respect to that of the NLO (no $K\Xi$) one is precisely due to the contribution of the set of $K\Xi$ data, with more disperse experimental points, rather than to a loss of accuracy in reproducing the measurements.
We observe that the inclusion of the NLO terms in the Lagrangian helps quite significatively in reducing the value of $\chi^2_{\rm d.o.f.}$ with respect to that obtained with the corresponding  WT fit at lowest order, especially when the $K\Xi$ data have been included. 
All the fits produce a quite stable value of $f$, lying very close to  $1.2f_\pi$. We observe that the WT fit, forced to accommodate the reproduction of the additional $K\Xi$ data set via unitarization loops, produces subtraction constants in the isospin $I=1$ channel, $a_{\pi\Lambda}$ and $a_{\eta\Sigma}$, which are one order of magnitude larger than what qualifies as being of ``natural" size \cite{OM}. The parameters obtained in the other fits presented in Table~\ref{tab2} are of reasonable size. It is found that, within about $2\sigma$ of their errors, the values of the subtraction constants and the $f$ parameter obtained in the NLO (no $K\Xi$)  and NLO fits are quite similar. However, the values of the low energy constants of the NLO Lagrangian ($b_0$, $b_D$, $b_F$ and $d_i$) obtained by the two fits show stronger differences. This means that these parameters are really sensitive to the data of the $K\Xi$ production reactions which should then be used to constrain their values, as done in the present work.
This is clearly reflected, not only in the results presented in Table~\ref{tab:thresh} and Fig.~\ref{fig2}, where we find a slight improvement in reproducing the threshold observables and the $K^- p\to K^- p,\, \bar{K}^0n, \pi^- \Sigma^+, \pi^+\Sigma^-, \pi^0 \Sigma^0, \pi^0\Lambda$ cross sections,  but also, and more especially, in the total cross section of the $K\Xi$ channels, which cannot be reproduced if the NLO terms are omitted. We can therefore conclude that the $K^- p \to K\Xi$ cross sections are crucial for constraining more precisely the low energy constants of the NLO Lagrangian.

Focusing now on the cascade production cross sections of Fig.~\ref{fig3}, we observe that the discrepancies between the NLO model and the data are larger in the vicinity of $2$~GeV and around $2.2$~GeV.  In the next section, we discuss an extension of the model that includes the presence of resonances explicitly to improve the description of the $K\Xi$ channels.  

\section{Inclusion of high spin hyperon resonances in the $\bar{K} N\rightarrow K\Xi$ transitions}
\label{reson}

The study shown above suggests the possibility to improve the description of data by implementing, in the $K\Xi$ channels, the contribution of resonances located around $2$~GeV and $2.2$~ GeV. This procedure is motivated by previous resonance models studying $\Xi$ production \cite{Nakayama:2006ty,Man:2011np,Sharov:2011xq,Shyam:2011ys,Jackson:2014hba}, which indicate the need to take into consideration the $\bar{K} N \rightarrow Y \rightarrow K \Xi$ transition amplitudes, where $Y$ stands for some high spin resonance coupling significantly to the $\bar{K} N$,  $K \Xi$  channels.

 In the energy range of interest, the PDG compilation \cite{PDG} gives eight resonances with three- and four-star status with masses lying in the range $1.89<M<2.35$ GeV, see Table~\ref{tab4}. Unfortunately, explicit branching ratios to $K\Xi$ decay have not been determined and only upper limits are given for two of these resonances: $<3\%$ for the $\Lambda(2100)$  and $<2\%$ for $\Sigma(2030)$. 
The natural main decay channels for all these resonances are $\pi\Lambda$ (for $\Sigma$ states), $\pi\Sigma$, and $\bar{K}N$, while the branching ratios to $K\Xi$ decay are expected to be small, since this process requires the creation of an additional $\bar{s}s$ pair. However, cross sections for the $\bar{K}N\to K\Xi$ reactions are more than two orders of magnitude smaller than, for example, those  of the $\bar{K}N\to \pi\Sigma$ and $\bar{K}N\to \bar{K}N$ processes, hence even small branching ratios can contribute appreciably to the former reactions. Thus, it is interesting to investigate the role of these above-threshold resonances. Note that most of these resonances have high spins, and therefore require a special treatment, analogous to that performed in \cite{Nakayama:2006ty,Man:2011np,Sharov:2011xq,Rush}.

\begin{table}
\begin{tabular}{ccccc}
\hline \\[-2.5mm]
 {Resonance}   & {\bf $I$ $(J^P)$}   & { Mass (MeV)}   &{ $\Gamma$ (MeV)}  & { $\Gamma_{K\Xi}/\Gamma $}   \\
\hline \\[-2.5mm]
 $\Lambda(1890)$ &	$0\left(\frac{3}{2}^+\right)$	& 1850 - 1910	& 60 - 200 & \\
 $\Lambda(2100)$ &	$0\left(\frac{7}{2}^-\right)$ & 2090 - 2110 &	100 - 250 & $< 3\%$ \\
 $\Lambda(2110)$ &	$0\left(\frac{5}{2}^+\right)$ & 2090 - 2140 &	150 - 250 &  \\
 $\Lambda(2350)$ &	$0\left(\frac{9}{2}^+\right)$ & 2340 - 2370 &	100 - 250 &  \\
 $\Sigma(1915)$ &	$1\left(\frac{5}{2}^+\right)$ & 1900 - 1935 &	80 - 160 &  \\
 $\Sigma(1940)$ &	$1\left(\frac{3}{2}^-\right)$ & 1900 - 1950 &	150 - 300 &  \\
 $\Sigma(2030)$ & $1\left(\frac{7}{2}^+\right)$ & 2025 - 2040 & 150 - 200 & $< 2\%$ \\
 $\Sigma(2250)$ &  $1\left(?^?\right)$  & 2210 - 2280 & 60 - 150 &  \\
\hline
\end{tabular}
\caption{Properties of the three- and four-star hyperon resonances in the mass range $1.89<M<2.35$ GeV taken from the results of the PDG review\cite{PDG}. }
\label{tab4}
\end{table}

Inspecting the resonance properties shown in Table~\ref{tab4} and the results of the NLO fit presented in Fig.~\ref{fig3}, the $\Sigma(2030)$ and $\Sigma(2250)$ resonances seem to be good candidates to be implemented in our model. The two selected candidates also coincide with the findings of Ref~ \cite{Sharov:2011xq}, where it was concluded that these two resonances gave the best account of data, after various combinations of several resonances from the eight known ones were examined. The spin and parity $J^\pi =7/2^+$ of the $\Sigma(2030)$ are well established. Those of the $\Sigma(2250)$ are not known,  but the most probable assignments are $5/2^-$ or $9/2^-$ \cite{PDG}. We choose $J^\pi =5/2^-$  to simplify the calculations, noting also that the $9/2^-$ choice does not change the results drastically as has been shown in \cite{Sharov:2011xq}.

\subsection{Formalism}
The $\bar{K} N \rightarrow K^+ \Xi^-$,  $K^0 \Xi^0$ reaction cross sections are obtained adding to the corresponding  chiral unitary model amplitude $T(s',s)$ described in the previous section, the contributions from the
$\bar{K} N\rightarrow \Sigma(2030) \rightarrow K\Xi$ and $\bar{K} N\rightarrow \Sigma(2250) \rightarrow K\Xi$ transitions, denoted by  $T^{{7/2}^+}(s',s)$ and $T^{{5/2}^-}(s',s)$ respectively, which are built as described below.

Adopting the Rarita-Schwinger method, as in \cite{Man:2011np}, the spin-5/2 and 7/2 baryon fields are described by a rank-2 tensor $Y_{5/2}^{\mu\nu}$  and a rank-3 tensor $Y_{7/2}^{\mu\nu\alpha}$, respectively. The Lagrangians are 
\begin{equation}
\mathcal{L}^{{5/2}^\pm}_{BYK}(q)= {\rm i}\frac{g_{BY_{5/2}K}}{m_K^2}\bar{B}\Gamma^{(\pm)}Y_{5/2}^{\mu\nu}\partial_\mu\partial_\nu K+H.c. \ ,
\label{L_Res5}
\end{equation}
for the spin-5/2 resonance and
\begin{equation}
\mathcal{L}^{{7/2}^\pm}_{BYK}(q)=-\frac{g_{BY_{7/2}K}}{m_K^3}\bar{B}\Gamma^{(\mp)}Y_{7/2}^{\mu\nu\alpha}\partial_\mu\partial_\nu \partial_\alpha K+H.c. \ ,
\label{L_Res7}
\end{equation}
for the spin-7/2 one, 
where $\Gamma^{(\pm)}= \binom{\gamma_5}{1}$, and $g_{BY_J K}$ stands for the baryon-kaon-$Y_{J}$ coupling.
The corresponding propagators are given by \cite{Man:2011np}:
\ba
S_{5/2}(q)&=&\frac{{\rm i}}{\slashed{q}-M_{Y_{5/2}}+{\rm i}\Gamma_{5/2}/2}\Delta^{\beta_1\beta_2}_{\alpha_1\alpha_2} ,\\
S_{7/2}(q)&=&\frac{{\rm i}}{\slashed{q}-M_{Y_{7/2}}+{\rm i}\Gamma_{7/2}/2}\Delta^{\beta_1\beta_2\beta_3}_{\alpha_1\alpha_2\alpha_3}, 
\label{S_Res}
\ea
where we have included the decay width, $\Gamma_{J}$, of the corresponding resonance. The tensors $\Delta$ are defined as: 
\ba
\Delta^{\beta_1\beta_2}_{\alpha_1\alpha_2} \left(\frac{5}{2} \right)&=&\frac{1}{2}\left(\theta^{\beta_1}_{\alpha_1}\theta^{\beta_2}_{\alpha_2}+ \theta^{\beta_2}_{\alpha_1}\theta^{\beta_1}_{\alpha_2}\right)
-\frac{1}{5}\theta_{\alpha_1\alpha_2}\theta^{\beta_1\beta_2} \no \\
& +& \frac{1}{10}\Big (\bar{\gamma}_{\alpha_1}\bar{\gamma}^{\beta_1}\theta^{\beta_2}_{\alpha_2} +\bar{\gamma}_{\alpha_1}\bar{\gamma}^{\beta_2}\theta^{\beta_1}_{\alpha_2} \no \\
&+&\,\bar{\gamma}_{\alpha_2}\bar{\gamma}^{\beta_1}\theta^{\beta_2}_{\alpha_1}+\bar{\gamma}_{\alpha_2}\bar{\gamma}^{\beta_2}\theta^{\beta_1}_{\alpha_1}\Big),
\label{Delta_5}
\ea 
\ba
\Delta^{\beta_1\beta_2\beta_3}_{\alpha_1\alpha_2\alpha_3} \left(\frac{7}{2} \right)&=&\frac{1}{36}\sum_{\scriptscriptstyle P(\alpha)P(\beta)}\Big(\theta^{\beta_1}_{\alpha_1}\theta^{\beta_2}_{\alpha_2}\theta^{\beta_3}_{\alpha_3} \no \\
&-&\frac{3}{7}\,\,\theta^{\beta_1}_{\alpha_1}\theta_{\alpha_2\alpha_3}\theta^{\beta_2\beta_3} \no \\
&-&\frac{3}{7}\,\,\bar{\gamma}_{\alpha_1}\bar{\gamma}^{\beta_1}\theta^{\beta_2}_{\alpha_2}\theta^{\beta_3}_{\alpha_3} \no \\
&+&\frac{3}{35}\,\bar{\gamma}_{\alpha_1}\bar{\gamma}^{\beta_1}\theta_{\alpha_2\alpha_3}\theta^{\beta_2\beta_3}\Big),
\label{Delta_7}
\ea 
where $\theta^{\nu}_{\mu}=g^{\nu}_{\mu}-q_\mu q^\nu/M_Y^2$ , $\bar{\gamma}_{\mu}=\gamma_{\mu}-q_\mu \slashed{q}/{M_Y^2}$, with $M_Y$ being the pertinent resonance mass. The tensor $\Delta$ for the spin-7/2 field, given in Eq. (\ref{Delta_7}), contains a summation over all possible permutations of Dirac indexes $\{\alpha_1 \alpha_2 \alpha_3\}$ and $\{\beta_1 \beta_2 \beta_3\}$.

From the Lagrangians of Eqs.~(\ref{L_Res5}) and (\ref{L_Res7}) one derives the baryon-kaon-$Y_{J}$ vertices: 
\ba
v^{{5/2}^\pm}_{BYK}&=&{\rm i}\frac{g_{BY_{5/2}K\phantom{\bar{K}}\!\!\!}}{m_K^2}k_\mu k_\nu \Gamma^{(\pm)} , \\ 
v^{{7/2}^\pm}_{BYK}&=&-\frac{g_{BY_{7/2}K\phantom{\bar{K}}\!\!\!}}{m_K^3}k_\mu k_\nu k_\sigma \Gamma^{(\mp)}  .
\label{Ver_Res}
\ea
The resonant contributions to the  $\bar{K} N\rightarrow K\Xi$ scattering amplitudes can then be obtained straightforwardly as:
\begin{equation} 
T_{\bar{K} N\rightarrow K\Xi}^{{5/2}^-}(s',s) = F_{5/2}\, \bar{u}_\Xi^{s'}(p')k'_{\beta_1}k'_{\beta_2}S_{5/2}(q)k^{\alpha_1}k^{\alpha_2}u_N^{s}(p) \ ,
 \label{T_5}
\end{equation}
and
$$
T_{\bar{K} N\rightarrow K\Xi}^{{7/2}^+}(s',s)=F_{7/2}\, \bar{u}_\Xi^{s'}(p')k'_{\beta_1}k'_{\beta_2}k'_{\beta_2}S_{7/2}(q)\,\cdot
$$
\begin{equation}
\cdot\, k^{\alpha_1}k^{\alpha_2}k^{\alpha_3} u_N^{s}(p)  \ ,
 \label{T_7}
\end{equation} 
where we have included a form factor:
\begin{equation} 
F_J=\frac{g_{\Xi Y_{J}K\phantom{\bar{K}}} \!\!\! g_{NY_{J}\bar{K}}}{m_K^{2J-1}}\exp\left(-\vec{k}^2/\Lambda^2_{J}\right)
\exp\left(-\vec{k'}^2/\Lambda^2_{J}\right)\ ,
\label{FF}
\end{equation} 
which inserts a phenomenological exponential function,
$\exp\left(-\vec{q\,}^2/\Lambda^2_{J}\right)$, in each vertex to suppress high powers of the meson momentum from the vertex contributions, see Eqs.~(\ref{T_5}),(\ref{T_7}), as it was done in \cite{Sharov:2011xq}. Strictly speaking the exponential factors in Eq.~(\ref{FF}) are not genuine form factors, since these should depend on the off-shell
momentum of the off-shell particle and should be normalized to $1$ at
the on-shell point. The ``form factor" in Eq.~(\ref{FF}) is just an ad-hoc function introduced to modify the
energy dependence of the resonance contribution. This prescription, however, is used in the resonance based model of \cite{Sharov:2011xq},  which inspired us to complement our study with the inclusion of resonances. So, we have decided to employ it for a more direct comparison with the above cited paper. Furthermore, in Ref.~\cite{Sharov:2011xq} the authors have studied different forms of form factor, and they claim that the $\exp\left(-\vec{q\,}^2/\Lambda^2_{J}\right)$ form gives the best $\chi^2_{\rm d.o.f.}$ result. In order to verify this statement we have also tried form factors depending on the four momentum squared of the off-shell resonance, either in the form $\exp\{-(k^2-M_{Y_{J}}^2)/\Lambda^2_{J}\}$, which has the same asymptotic behavior at high values of the meson tri-momentum $\vec{q}$, or via the function $\Lambda^4_J/[\Lambda^4_J+(k^2-M_{Y_{J}}^2)^2]$, employed in the recent work of \cite{jackson}. In the results sections we will discuss the consequences of the choice of form factor on the data fitting.

Finally, for the initial $K^- p$, $\bar{K}^0 n$ channels and final $K^+ \Xi^-$, $K^0 \Xi^0$ ones we obtain
$$
\sqrt{4M_pM_\Xi} T^{\rm tot}_{ij}(s',s) = \sqrt{4M_pM_\Xi} T_{ij}(s',s)
$$
\begin{equation}
+\, T_{ij}^{{5/2}^-}(s',s)\, +\, T_{ij}^{{7/2}^+}(s',s)\ ,
\label{T_reso}
\end{equation}
where the amplitudes $ T_{ij}^{R}(s',s)$ contain the appropiate Clebsh-Gordan coefficients projecting the states $i$ and $j$ states into the isospin 1 of the $5/2^-$ and $7/2^+$ resonances included here. One can then proceed to derive the observables, following Eqs. (\ref{M_matrix})-(\ref{branch_ratios}). 

The chiral unitary model of the previous section is limited to $s$-wave interactions and, therefore, gives rise to flat differential cross sections. On the contrary,  the high spin resonance mechanisms described in this section introduce an angular dependence in the amplitudes of the $K\Xi$ production channels, permitting a study of the differential cross sections for these channels, which are given by 
\begin{equation}
\frac{d\sigma_{ij}}{d\Omega}=\frac{1}{64\pi^2}\frac{4M_iM_j}{s}\frac{k_j}{k_i}S_{ij} ,
 \label{dsigma_0}
\end{equation}
where $S_{ij} $ is obtained from Eq.~(\ref{M_matrix}), but employing the $T^{\rm tot}_{ij}$ amplitude of Eq.~(\ref{T_reso}).

\begin{table*}[ht]
\begin{tabular}{lcccccc}
\hline \\[-2.5mm]
&{$\gamma$} & {$R_n$} & {$R_c$} & {$a_p(K^-p \rightarrow K^- p)$} & {$\Delta E_{1s}$} & {$\Gamma_{1s}$} \\
\hline \\[-2.5mm]
NLO*    & $2.37$   &  $0.189$   &  $0.664$ & $-0.69+{\rm i\,}0.86$ & $300$ & $570$   \\%
WT+RES  & $2.37$   &  $0.193$   &  $0.667$ & $-0.73+{\rm i\,}0.81$ & $307$ & $528$   \\%
NLO+RES & $2.39$   &  $0.187$   &  $0.668$ & $-0.66+{\rm i\,}0.84$ & $286$ & $562$   \\%
\hline \\[-2.5mm]
Exp.    &	$2.36$     &	$0.189$      &   $0.664$   & $ -0.66 + {\rm i\,}0.81$ & $283$ & $541$ \\
        &  $\pm 0.04$  &    $\pm 0.015$  & $\pm 0.011$ & $(\pm0.07)+ {\rm i\,}(\pm0.15)$ &	$\pm36$ & $\pm92$\\
\hline
\end{tabular}
\caption{Threshold observables obtained from the  NLO*, WT+RES and NLO+RES fits  explained in the text. Experimental data is taken from cite{}.}
\label{tab5}
\end{table*}

\subsection{Data treatment and fits}

Since the new high spin resonant terms produce angular dependent scattering amplitudes, we will consider, in addition to the total cross sections and threshold observables listed in Table~\ref{tab1},  the differential cross sections of the  $K^-p \to K\Xi$ reactions taken from the same sources \cite{exp_5, exp_6, exp_7, exp_8, exp_9, exp_10, exp_11}. More specifically, the fits in this section will include 2 new observables: the 235 differential cross section points from the $K^+ \Xi^-$ production reaction and 76 differential cross section points from the $K^0 \Xi^0$ one. Thus we increase the total number of experimental points to 477 instead of the 161 employed in the fits of the previous section. With the aim of preserving the same weight for each observable, the same definition of the $\chi^2_{\rm d.o.f.}$, Eq.~(\ref{Chi^2_dof}), is employed. However, in the new fit the overall weight of the $K\Xi$ channels is larger, since there are two new observables related to these.  

It must be also mentioned that large amount of new points, more dispersed, could rise the contribution to $\chi^2_{\rm d.o.f.}$, but, as we will see, we gain in having a better overall description of the $K^- p \to K\Xi$ reactions while fully respecting an acceptable accuracy for the other observables.

We will present results for three different fits, all of them employing the data of the previous section plus the differential cross section data of the $K\Xi$ production reactions: 

i) A fit denoted by NLO*, which employs the NLO interaction kernel without any additional resonance contribution. Thus, this fit is completely analogous to the NLO fit from the previous session, and correspondingly the resulting curves for the NLO* differential cross sections of the  $K^-p \to K\Xi$ reactions will be flat, without any angular dependence. However taking into account the new experimental points of the differential cross sections we give a larger weight to the $K\Xi$ channels, as discussed above, therefore we expect a slight modification of the model parameters with respect to NLO fit from previous section. We would like to remind that there are 14 free parameters involved in NLO* fit: the pion decay constant $f$, the six subtraction constants, and the seven low energy constants of the NLO Lagrangian.

\begin{table*}
\begin{tabular}{lrrr}
\hline \\[-2.5mm]
                                                  &   {NLO*}      & {WT+RES}       & {NLO+RES}  \\
\hline \\[-2.5mm]
$a_{\bar{K}N} \ (10^{-3})$                   & $6.799\pm0.701$   & $-1.965\pm2.219$   & $6.157\pm0.090$  \\
$a_{\pi\Lambda}\ (10^{-3})$                       & $50.93\pm9.18$  & $-188.2\pm131.7$ & $59.10\pm3.01$ \\
$a_{\pi\Sigma}\ (10^{-3})$                        & $-3.167\pm1.978$  & $0.228\pm2.949$    & $-1.172\pm0.296$  \\
$a_{\eta\Lambda}\ (10^{-3})$                      & $-15.16\pm12.32$ & $1.608\pm2.603$   & $-6.987\pm0.381$  \\
$a_{\eta\Sigma}\ (10^{-3})$                       & $-5.325\pm0.111$  & $208.9\pm151.1$  & $-5.791\pm0.034$  \\
$a_{K\Xi}\ (10^{-3})$                             & $31.00\pm9.441$  & $43.04\pm25.84$  & $32.60\pm11.65$ \\
$f/f_\pi$                                         & $1.197\pm0.011$   & $1.203\pm0.023$  & $1.193\pm0.003$   \\
$b_0$ (GeV$^{-1}$)                               & $-1.158\pm0.021$  &      \multicolumn{1}{c}{-}              & $-0.907\pm0.004$  \\
$b_D$ (GeV$^{-1}$)                                & $0.082\pm0.050$   &      \multicolumn{1}{c}{-}              & $-0.151\pm0.008$  \\
$b_F$ (GeV$^{-1}$)                                & $0.294\pm0.149$   &      \multicolumn{1}{c}{-}              & $0.535\pm0.047$   \\
$d_1$ (GeV$^{-1}$)                                & $-0.071\pm0.069$  &      \multicolumn{1}{c}{-}              & $-0.055\pm0.055$  \\
$d_2$ (GeV$^{-1}$)                                & $0.634\pm0.023$   &      \multicolumn{1}{c}{-}              & $0.383\pm0.014$   \\
$d_3$ (GeV$^{-1}$)                                & $2.819\pm0.058$   &      \multicolumn{1}{c}{-}              & $2.180\pm0.011$   \\
$d_4$ (GeV$^{-1}$)                                & $-2.036\pm0.035$  &      \multicolumn{1}{c}{-}              & $-1.429\pm0.006$  \\
$g_{\Xi Y_{5/2}K\phantom{\bar{K}}} \!\!\!\cdot  g_{NY_{5/2}\bar{K}}$ &
    \multicolumn{1}{c}{-}           & $-5.42\pm15.96$  & $8.82\pm5.72$  \\
$g_{\Xi Y_{7/2}K\phantom{\bar{K}}} \!\!\! \cdot g_{NY_{7/2}\bar{K}}$ &
    \multicolumn{1}{c}{-}          & $-0.61\pm14.12$   & $0.06\pm0.20$  \\
$\Lambda_{5/2}$ (MeV)                             &   \multicolumn{1}{c}{-}        & $576.7\pm275.2$  & $522.7\pm43.8$ \\
$\Lambda_{7/2}$ (MeV)                             &        \multicolumn{1}{c}{-}          & $623.7\pm287.5$  & $999.0\pm288.0$ \\
$M_{Y_{5/2}}$ (MeV)                               &        \multicolumn{1}{c}{-}           & $2210.0\pm39.8$  & $2278.8\pm67.4$   \\
$M_{Y_{7/2}}$ (MeV)                               &        \multicolumn{1}{c}{-}           & $2025.0\pm9.4$    & $2040.0\pm9.4$    \\
$\Gamma_{5/2}$ (MeV)                              &        \multicolumn{1}{c}{-}           & $150.0\pm71.3$   & $150.0\pm54.4$     \\
$\Gamma_{7/2}$ (MeV)                              &        \multicolumn{1}{c}{-}           & $200.0\pm44.6$   & $200.0\pm32.3$     \\
\hline \\[-2.5mm]
$\chi^2_{\rm d.o.f.}$                                   & $1.48$            & $2.26$             &      $1.05$           \\
\hline
\end{tabular}
\caption{Values of the parameters and the corresponding  $\chi^2_{\rm d.o.f.}$, defined as in Eq.~(\ref{Chi^2_dof}), for the different fits described in the text. The value of the pion decay constant is $f_{\pi}=93$ MeV and the subtraction constants are taken at a regularization scale $\mu=1$ GeV.}
\label{tab6}
\end{table*}

ii) Another fit, denoted by WT+RES, which employs the lowest order kernel of the chiral Lagrangian and adds the resonant terms described in this section. This fit has 15 free parameters: the same seven parameters as those for the lowest order fits of the previous section ($f$ and the 6 subtraction constants) plus eight new parameters associated to the resonant terms, namely masses and widths of the resonances ($M_{Y_{5/2}}$, $M_{Y_{7/2}}$, $\Gamma_{5/2}$ and $\Gamma_{7/2}$), the product of couplings  ($g_{ \Xi Y_{5/2}K\phantom{\bar{K}}\!\!\!}\cdot g_{NY_{5/2}\bar{K}}$ and $g_{ \Xi Y_{7/2}K\phantom{\bar{K}}\!\!\!}\cdot g_{NY_{7/2}\bar{K}}$) and the cut-off in the form factors ($\Lambda_{5/2}$ and $\Lambda_{7/2}$). This fit aims at exploring whether the background terms could be accounted only through the lowest order chiral Lagrangian, while the $K\Xi$ channels can be covered by the resonant terms.

\begin{figure*}[!ht]
\centering
 \includegraphics[width=4.5in]{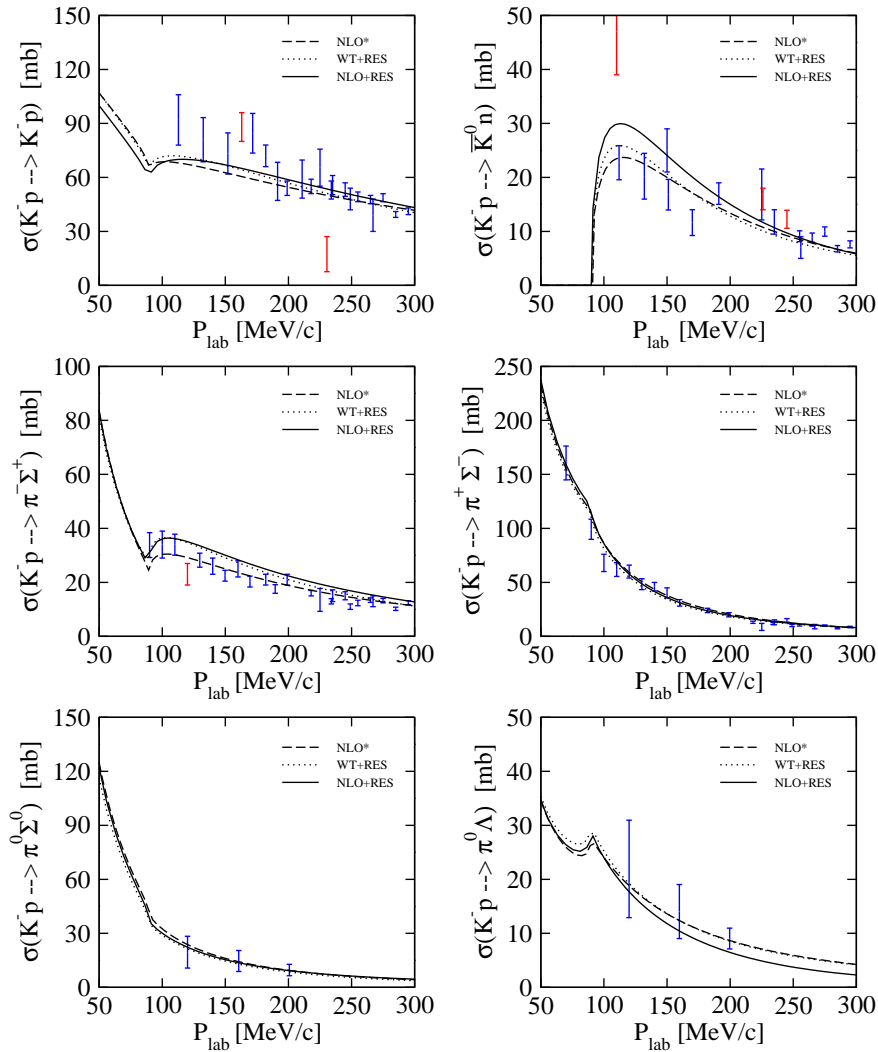}
\caption{Total cross sections of the $K^- p\to K^- p, \bar{K}^0n, \pi^- \Sigma^+, \pi^+\Sigma^-, \pi^0 \Sigma^0, \pi^0\Lambda $ reactions for the NLO* fit (dashed line), the WT+RES fit (dotted line) and the NLO+RES fit (solid line). Experimental data are from \cite{exp_1,exp_2, exp_3, exp_4}. The points in red have not been included in the fitting procedure.} 
  \label{fig4}
\end{figure*}

\begin{figure}[!ht]
\centering
 \includegraphics[width=2.8 in]{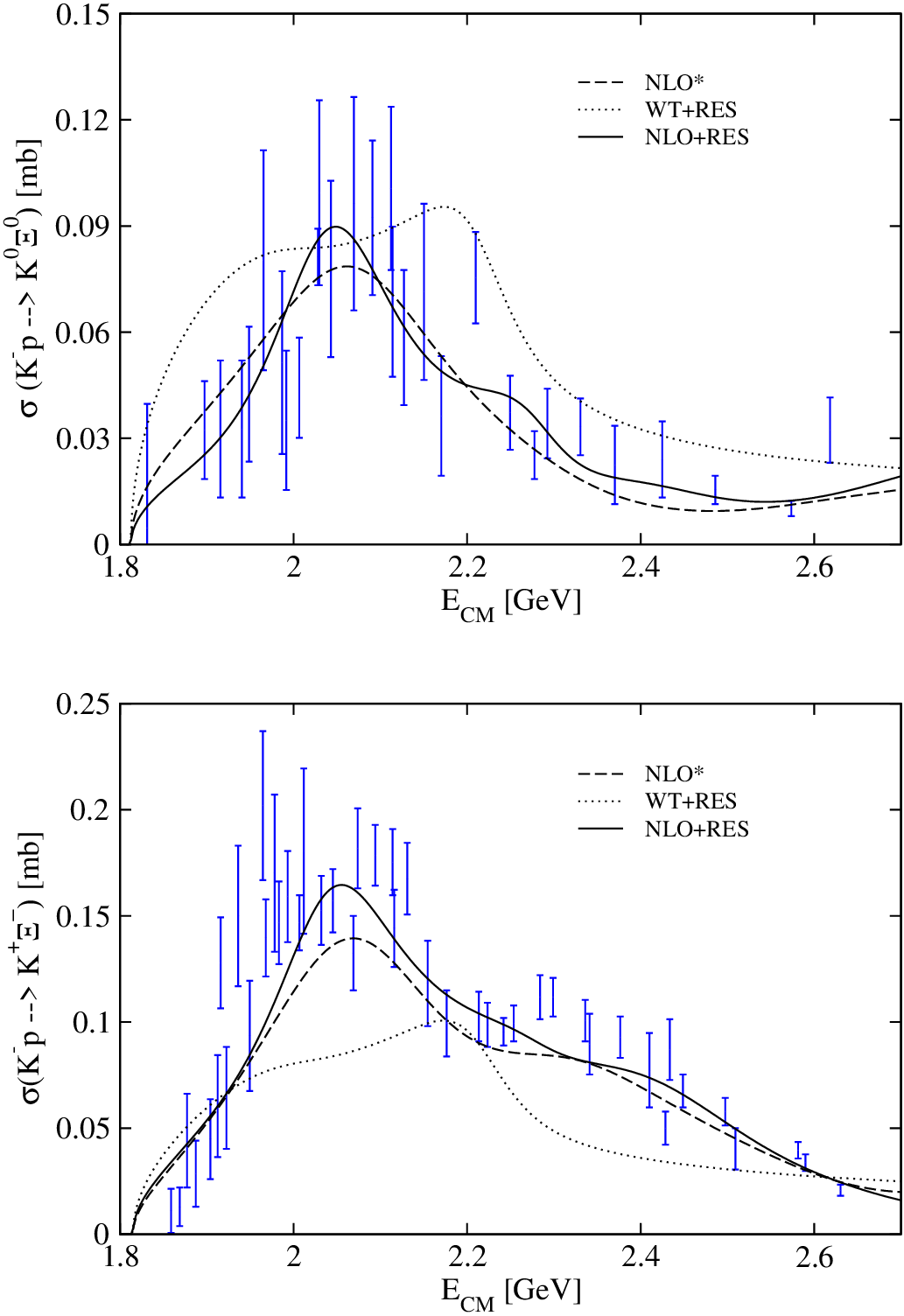}
\caption{Total cross sections of the $K^- p\to K^0 \Xi^0, K^+ \Xi^-$ reactions for the NLO* fit (dashed line), the WT+RES fit (dotted line) and the NLO+RES fit (solid line), see the text for more details. Experimental data are from \cite{exp_5, exp_6, exp_7, exp_8, exp_9, exp_10, exp_11}.} 
  \label{fig5}
\end{figure}

iii) Finally, a  fit denoted by NLO+RES, which incorporates the NLO interaction kernel together with the high spin resonance contributions in the $K^- p \to K^+ \Xi^-$, $K^0 \Xi^0$ channels. This fit determines 22 free parameters: the same fourteen as in the NLO* fit and the new eight parameters associated to the resonant terms. This is the most complete calculation that, upon comparison with the results of the previous WT+RES fit, will assess the actual role of the NLO terms in the chiral Lagrangian and will determine the value of their low energy constants.

We note that not all parameters are fully free. We constrain masses and widths of the resonances to lie within the ranges given in the PDG compilation \cite{PDG} (see table \ref{tab4}) and the form-factor cut-off values are constrained between 500 MeV$<\Lambda_{J}<1000$ MeV.

\subsection{Results and discussion}
\label{subsec:results2}

In this section we discuss  the results of the fits described above which, differently to those shown in Sect.~\ref{subsec:results1}, have also included the differential $K^-p \to K\Xi$ cross sections in the fitting procedure. The results for the threshold observables shown in Table~\ref{tab6} indicate that, even if the fits now adjust new data at higher energies and may contain the additional effect of resonant terms, as in the case of  WT+RES and NLO+RES, the low energy data keeps being very well described. A similar situation is found when inspecting the cross sections obtained from the three fits for the $K^- p\to K^- p,\, \bar{K}^0n, \pi^- \Sigma^+, \pi^+\Sigma^-, \pi^0 \Sigma^0, \pi^0\Lambda$ reactions shown in figure \ref{fig4}.

\begin{figure*}[!htb]
\centering
 \includegraphics[width=4.5in]{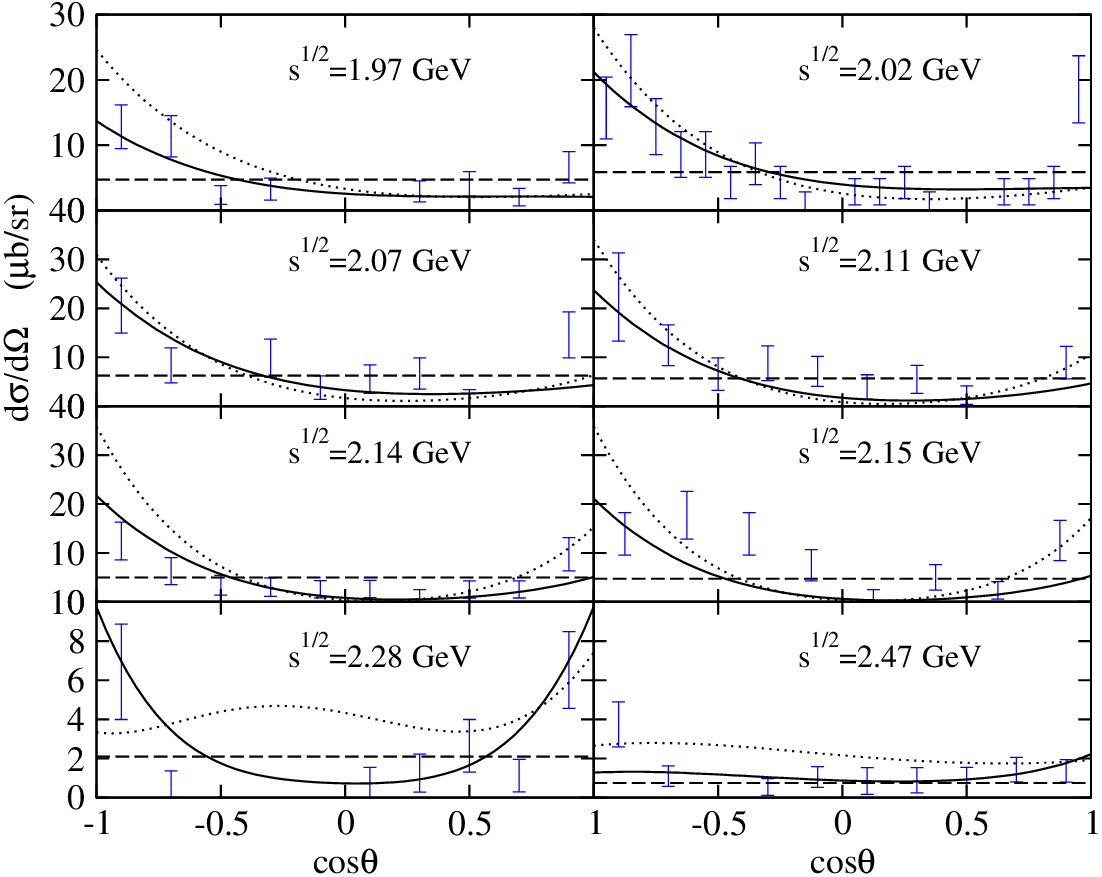}
\caption{Differential cross section of the $K^- p\to K^0 \Xi^0$ reaction for the NLO* fit (dashed line), the WT+RES fit (dotted line) and the NLO+RES fit (solid line), see the text for more details. Experimental data are from \cite{exp_5, exp_6, exp_7, exp_8, exp_9, exp_10, exp_11}.} 
  \label{fig6}
\end{figure*}
\begin{figure*}[ht]
\centering
 \includegraphics[width=4.5in]{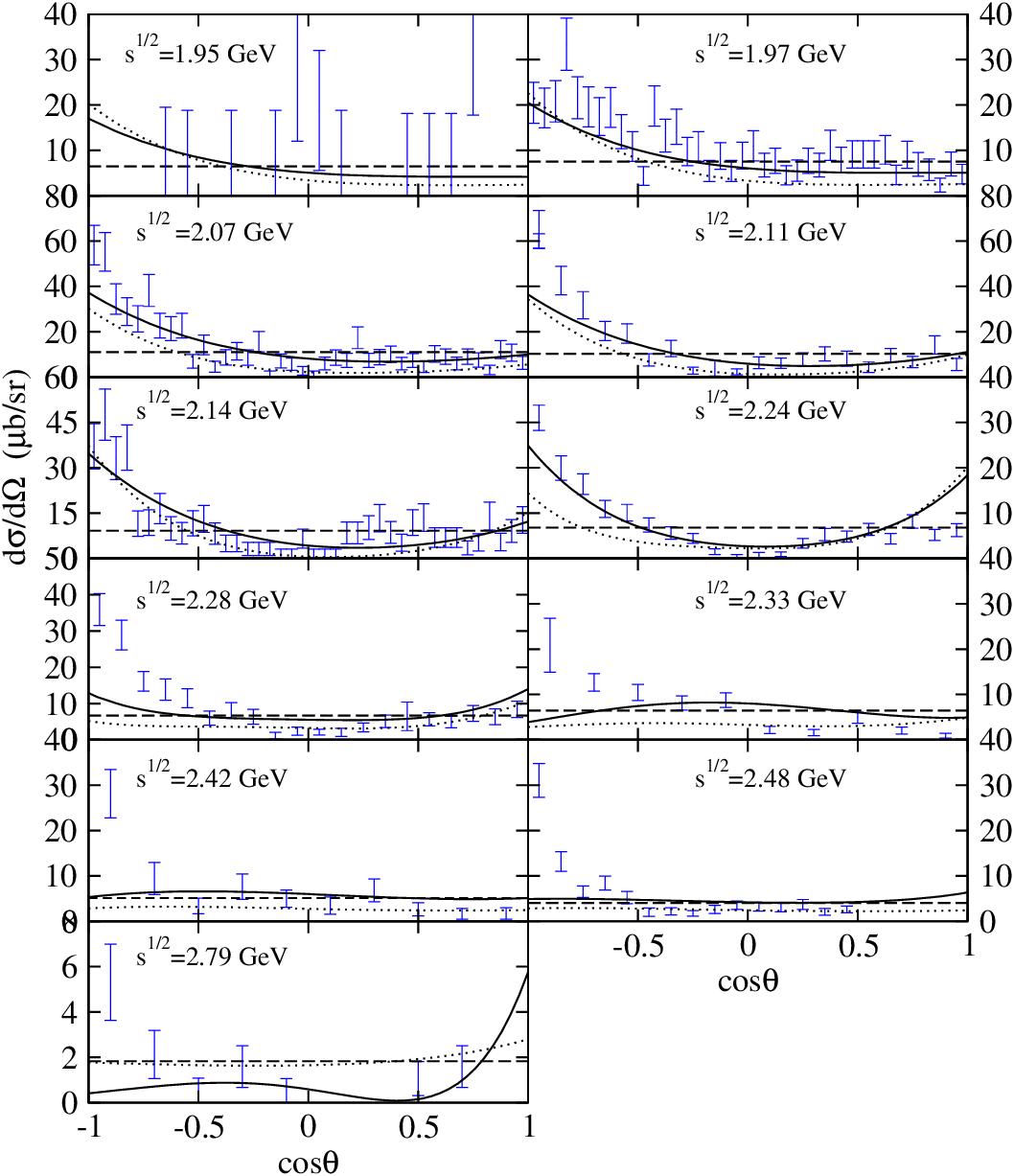}
\caption{Differential cross section of the $K^- p\to K^+ \Xi^-$ reaction for the NLO* fit (dashed line), the WT+RES fit (dotted line) and the NLO+RES fit (solid line), see the text for more details. Experimental data are from \cite{exp_5, exp_6, exp_7, exp_8, exp_9, exp_10, exp_11}.} 
  \label{fig7}
\end{figure*}

Obviously, the differences between these fits are more evident in the total and differential cross sections of the $K\Xi$ production channels shown in
Figs.~\ref{fig5}, \ref{fig6} and \ref{fig7}.
First we note that the total cross sections for $K\Xi$ production obtained from the NLO* fit (dashed lines in Fig.~\ref{fig5}) are in reasonable agreement with the data, even if the resonant terms are not included. As it was discussed above, this NLO* fit is very similar to the NLO one of the previous section, but it also tries to accommodate the differential $K\Xi$ production cross section data,  which can only be adjusted on average, as shown by the dashed lines in Figs.~\ref{fig6} and \ref{fig7}, because of the flat distribution characteristic of $s$-wave models. 

In order to account for some structure in the differential $K\Xi$ production cross sections we need to implement the resonant terms. When they are added to the unitarized amplitudes obtained from the lowest order chiral Lagrangian, one finds the results denoted by the dotted lines, or WT+RES fit, in  Figs.~\ref{fig5}, \ref{fig6} and \ref{fig7}. It is clear that, although some structure is gained in the differential cross sections and, hence, their description improves substantially than in the absence of resonances, the total $K\Xi$ production cross sections are poorly reproduced by the WT+RES fit.  In other words, the background terms encoded in the lowest order chiral Lagrangian, which only contribute via unitarization, are insufficient to account for the whole set of $K\Xi$ production data satisfactorily. This situation is remedied when the chiral Lagrangian is taken at NLO. In this case, one finds a clear overall improvement in the description of the data. The solid lines in Figs.~\ref{fig5}, \ref{fig6} and \ref{fig7} clearly demonstrate that the NLO+RES fit reproduces satisfactorily the $K\Xi$ total cross sections, while accounting  quite reasonably for the differential ones. Our model fails especially at backward angles for the higher $K^-$ energies. Obviously, including the role of additional hyperon resonances in s- and u-channel configurations could improve these deficiencies. However, this goes beyond the purpose of this paper, which focuses on demonstrating the essential role that the $K^- p \to K \Xi$ reactions have in determining of the low energy constants of the NLO chiral Lagrangian, as we emphasize again below. It is also worth mentioning that the inclusion of the high-spin resonances in the fit is very time consuming: the calculations are prolonged by factor 100, from several hours to several weeks.

One can judge the goodness of the fits discussed  in this section by inspecting the obtained $\chi^2_{\rm d.o.f.}$, shown in Table~\ref{tab6} together with the values of the fitted parameters.
The first observation that we can make is that, even if the NLO* fit shows a similar quality as the NLO fit of the previous section in reproducing the cross section data, it has twice its  $\chi^2_{\rm d.o.f.}$ value. This is due to the additional differential cross section data employed in the NLO* fit, which can only be reproduced on average, leaving the predictions quite far away from the experimental points in some cases.  Also we can see that the parameters of these two fits are rather similar.

It is interesting to point out that, although the resonant terms naturally improve the description of the  $K\Xi$ differential cross section data, when the chiral Lagrangian is kept up to the lowest order, then  the corresponding  WT+RES $\chi^2_{\rm d.o.f.}$ value increases in about one unit with respect to the non-resonant NLO* fit. This just reflects the inability of the lowest order Lagrangian of producing enough strength, which we recall comes from unitarization, to interfere efficiently with that of the resonant terms. This gives rise to a poor description of the $K\Xi$ total cross section data  and, consequently, to an unreasonably large $\chi^2_{\rm d.o.f.}$ value. As in the previous section, the size of some of the subtraction constants of this fit turn out to be unphysically large. We then find again that the NLO terms of the chiral Lagrangian are essential to account for the $K\Xi$ data. This is reflected in a reduction of the corresponding NLO+RES  $\chi^2_{\rm d.o.f.}$ value, which turns out to be of around one.

We have also performed fits with the two choices of form factor that depend on the off-shell four-momentum of the resonance and are normalized to 1 at the on-shell point, namely  $\exp\{-(k^2-M_{Y_{J}}^2)/\Lambda^2_{J}\}$ and $\Lambda^4_J/[\Lambda^4_J+(k^2-M_{Y_{J}}^2)^2]$ (see discussion after Eq. (\ref{FF})).
We have found that the  $\chi^2_{\rm d.o.f.}$ worsens, giving in both cases a value of 1.25 versus the 1.05 value obtained for the ad-hoc prescription, in complete agreement to the claims made in Ref.~\cite{Sharov:2011xq}. Interestingly, the corresponding NLO parameters do not change significantly and remain quite similar to the NLO+RES ones shown in Table~\ref{tab6}.

The important role of the $K\Xi$ channels in constraining the NLO terms of the chiral Lagrangian has already been
shown in the previous section, where the corresponding low energy constants, obtained including the $K\Xi$ production total cross section data in the NLO fit, changed appreciably with respect to those of the NLO (no $K\Xi$) fit.  In this section, we have seen how the description of data, which now includes the additional $K\Xi$ differential cross sections, is further improved when we supplement the NLO Lagrangian with the resonant terms. 
We observe that, although there is a slight readjustment of the parameters of the NLO+RES fit with respect to those of the NLO* fit, they have gained in precision significantly. This is due to the stabilizing role of the resonant terms, which implement an important part of the energy dependencies, hence relegating the role of the NLO Lagrangian contribution to be a smooth background.  This is in line to the contribution of the contact term introduced ad hoc in the resonant model of Ref.~\cite{Jackson:2014hba} to account for the strong $\Xi$ production data.

We also comment on the resonance parameters obtained by our NLO+RES fit. First of all, we would like to remind the reader that the masses and widths are constrained to lie within the experimentally measured bounds \cite{PDG}. As we can see in Table \ref{tab6} the product of couplings and the form factors are not very well constrained by the fit. 

As mentioned already, we complemented our study with the inclusion of high spin hyperonic resonances being inspired by the work of \cite{Sharov:2011xq}, but we would like to point out that a direct comparion of the resonance parameters of our model with the those of \cite{Sharov:2011xq} is not straightforward. This is also the case when comparing similar resonance based models. For instance,  the resonance parameters obtained in \cite{Sharov:2011xq} are quite different than those in \cite{jackson}\footnote{For a proper comparison, note that the dimensionless couplings given in \cite{jackson}, as well as those of the present work,  are given in units of the kaon mass, while those of \cite{Sharov:2011xq} use the pion mass.}, and the high-spin resonance contributions may differ by more than a factor of two in both resonance models. The reason is that the effect of these resonances depends very much on the interference with the background terms. Clearly, different backgrounds will result in rather different coupling sizes and even signs, as it was shown in \cite{jackson}. However, the big advantage of our approach is that our "background terms" are completely determined by a theoretically supported chiral model.  
Still, trying to compare our results with those of \cite{Sharov:2011xq}, where``form factors" of the same type have been used, 
we observe that, while our value of $g_{\Xi Y_{5/2}K\phantom{\bar{K}}} \!\!\!\cdot  g_{NY_{5/2}\bar{K}}$ turns to be comparable, although having an opposite sign, to that obtained in the resonant model of Ref.~\cite{Sharov:2011xq}, the product $g_{\Xi Y_{7/2}K\phantom{\bar{K}}} \!\!\!\cdot  g_{NY_{7/2}\bar{K}}$ is almost three orders of magnitude smaller. 
Note, however, that this has also to be viewed together with the effect of the form-factor, which in the present work is more moderate, since the  cut-off values turn out to be larger, especially for the $7/2^+$ resonance, than the $440$ MeV value employed in \cite{Sharov:2011xq}.

We have also tried to make a fit with 3 resonances, implementing an additional P-wave state in our model, lying close to
the $K \Xi$ threshold. This could be for example the $\Lambda(1890)$ $3/2^-$ resonance, also included in \cite{jackson}. However, we find that a resonance of this type does not improve substantially  
 the quality of the fit. The change of $\chi^2_{\rm d.o.f.}$ from 1.05 to 1.04, while keeping the NLO parameters rather stable and similar to those quoted in Table~\ref{tab6}, does not compensate, in our opinion, the increase of complexity of the problem and of the necessary computing time.

Finally, we would like to mention again the (unexpected) stability of the pion decay width parameter $f$ which stays around $1.195$ in all the fits.

\section{Conclusions}
\label{conclusions}

In this work, we have presented a study of the $S=-1$ meson-baryon interaction, employing a chiral SU(3) Lagrangian up to next-to-leading order and implementing unitarization in coupled channels. The parameters of the Lagrangian have been fitted to a large set of experimental scattering data in different two-body channels, to $\gamma$, $R_n$ and $R_c$ branching ratios, and to the precise SIDDHARTA value of the energy shift and width of kaonic hidrogen. In contrast to  other works, we have also constrained our model to reproduce the $K^- p\to K^+\Xi^-, K^0\Xi^0$ reactions, since they  become  especially sensitive to the NLO terms, as they cannot proceed with the LO Lagrangian, except indirectly via unitarization contributions.

By comparing different fitting procedures, 
we have shown in the first part of our study that the NLO order terms of the chiral Lagrangian are important and a necessary ingredient of the model, since they help in achieving a better description of data.  A novelty of the present work is that we have clearly established
the sensitivity of the NLO Lagrangian to
the  $K^- p \to K\Xi$ reactions. Therefore, by implementing the cross section data  for $K\Xi$ production in the fitting procedure, we have been able to obtain more accurate values of the low energy constants of the NLO chiral Lagrangian.

In the second part of this work, we have allowed for the explicit contribution  of two high spin hyperon resonances to the $K^- p \to K\Xi$ amplitudes, aiming at establishing an appropriate amount for the background, which in this work is associated to the chiral contributions, and, hence, obtain more reliable values of the associated low energy constants. Since the resonant terms introduce an angular dependence in the amplitudes, we also attempt the description of the $K\Xi$ differential cross sections. We find the resonant terms to have a double benefit. On the one hand, they allow for a reasonable overall description of the scattering data, including the total and the differential cross sections of the $K\Xi$ production reactions. On the other hand, by absorbing certain structures of the cross section, the inclusion of resonant contributions permit finding a more stable solution and therefore more precise values of the low energy constants of the chiral unitary model.

\begin{table*}
\begin{tabular}{|l|cccccccccc|}
\hline
 & & & & & & & & & & \\[-2.5mm]
  &{\bf $K^-p$}&{\bf $\bar{K}^0n$}&{\bf $\pi^0\Lambda$}& {\bf $\pi^0\Sigma^0$}&{\bf $\eta\Lambda$}&{\bf $\eta\Sigma^0$}&{\bf $\pi^+\Sigma^-$}&{\bf $\pi^-\Sigma^+$}&{\bf $K^+\Xi^-$}&{\bf $K^0\Xi^0$}  \\
\hline
 & & & & & & & & & & \\[-2.5mm]
{\bf $K^-p$}         & $2$ & $1$ & $\sqrt{3}/2$  & $1/2$ & $3/2$ & $\sqrt{3}/2$  & $0$ & $1$ & $0$           & $0$     \\
{\bf $\bar{K}^0n$}   &     & $2$ & $-\sqrt{3}/2$ & $1/2$ & $3/2$ & $-\sqrt{3}/2$ & $1$ & $0$ & $0$           & $0$     \\
{\bf $\pi^0\Lambda$} &     &     &    $0$        &  $0$  &  $0$  &  $0$          & $0$ & $0$ & $\sqrt{3}/2$  & $-\sqrt{3}/2$ \\
{\bf $\pi^0\Sigma^0$}&     &     &               &  $0$  &  $0$  &  $0$          & $2$ & $2$ &  $1/2$        & $1/2$ \\
{\bf $\eta\Lambda$}  &     &     &               &       &  $0$  &  $0$          & $0$ & $0$ &  $3/2$        & $3/2$     \\             
{\bf $\eta\Sigma^0$} &     &     &               &       &       &  $0$          & $0$ & $0$ &  $\sqrt{3}/2$ & $-\sqrt{3}/2$ \\             
{\bf $\pi^+\Sigma^-$}&     &     &               &       &       &               & $2$ & $0$ &  $1$          & $0$ \\
{\bf $\pi^-\Sigma^+$}&     &     &               &       &       &               &     & $2$ &  $0$          & $1$ \\
{\bf $K^+\Xi^-$}     &     &     &               &       &       &               &     &     &  $2$          & $1$ \\
{\bf $K^0\Xi^0$}     &     &     &               &       &       &               &     &     &               & $2$ \\

\hline

\end{tabular}
\caption{$C_{ij}$ coefficients of Eq. (\ref{WT}).}
\label{tab7}
\end{table*}

Summarizing, either taking into account or not the high spin hyperon resonances, the present work has clearly shown for the first time that the NLO corrections of the chiral Lagrangian are absolutely necessary to reproduce the  $K^- p\to K \Xi$ reaction data, and, conversely, taking into account these data permits a more precise and trustable determination of the corresponding NLO parameters.

\begin{acknowledgments}
We would like to thank D. A. Sharov and D. E. Lanskoy their kindness for making available to us their compilation of the large set of $K\Xi$ production data.
This work is supported  by the European Community - Research Infrastructure Integrating Activity Study of Strongly Interacting Matter (HadronPhysics3, Grant Agreement Nr 283286) under the 7th Framework Programme, by the contract FIS2011-24154 from MICINN (Spain), and by the Ge\-ne\-ra\-li\-tat de Catalunya, contract 2014SGR-401.
\end{acknowledgments}

\appendix 
\section{Tables of coefficients}
\label{appendix}
Table \ref{tab7} presents the $C_{ij}$ coefficients of Eq. (\ref{WT}), while Table \ref{tab8} presents the $D_{ij}$, $L_{ij}$ coefficients of Eq.~(\ref{V_NLO}).

%\newpage

\begin{table*}[h]
\begin{sideways}
\begin{tabular}{|l|cccccccccc|}
\multicolumn{11}{c}{{\rm $D_{ij}$ coefficients}} \\[2.5mm]
\hline
 & & & & & & & & & & \\[-2.5mm]
  &{\bf $K^-p$}&{\bf $\bar{K}^0n$}&{\bf $\pi^0\Lambda$}& {\bf $\pi^0\Sigma^0$}&{\bf $\eta\Lambda$}&{\bf $\eta\Sigma^0$}&{\bf $\pi^+\Sigma^-$}&{\bf $\pi^-\Sigma^+$}&{\bf $K^+\Xi^-$}&{\bf $K^0\Xi^0$}  \\
 \hline
& & & & & & & & & & \\[-2.5mm]
{\bf $K^-p$} &$4(b_0+b_D)m_K^2$ & $2(b_D+b_F)m_K^2$ & $\frac{-(b_D+3b_F)\mu_1^2}{2\sqrt{3}}$ & $\frac{(b_D-b_F)\mu_1^2}{2}$ &     $\frac{(b_D+3b_F)\mu_2^2}{6}$ & $\frac{-(b_D-b_F)\mu_2^2}{2\sqrt{3}}$ & $0$ & $(b_D-b_F)\mu_1^2$ & $0$ & $0$ \\ [2mm]
{\bf $\bar{K}^0n$}   &                 &$4(b_0+b_D)m_K^2$&$\frac{(b_D+3b_F)\mu_1^2}{2\sqrt{3}}$ &$\frac{(b_D-b_F)\mu_1^2}{2}$&   $\frac{(b_D+3b_F)\mu_2^2}{6}$ & $\frac{(b_D-b_F)\mu_2^2}{2\sqrt{3}}$  & $(b_D-b_F)\mu_1^2$  & $0$ & $0$ &  $0$    \\ [2mm]
{\bf $\pi^0\Lambda$} &                 &                 &$\frac{4(3b_0+b_D)m_{\pi}^2}{3}$      & $0$                        &    $0$ & $\frac{4b_Dm_{\pi}^2}{3}$ & $0$ & $0$ & $\frac{-(b_D-3b_F)\mu_1^2}{2\sqrt{3}}$ & $\frac{(b_D-3b_F)\mu_1^2}{2\sqrt{3}}$ \\ [2mm]
{\bf $\pi^0\Sigma^0$}&                 &                 &                                      & $4(b_0+b_D)m_{\pi}^2$      &    $\frac{4b_Dm_{\pi}^2}{3}$ & $0$ & $0$ & $0$ & $\frac{(b_D+b_F)\mu_1^2}{2}$ & $\frac{(b_D+b_F)\mu_1^2}{2}$    \\ [2mm]
{\bf $\eta\Lambda$}  &                 &                 &                                      &                            &           $\frac{4(3b_0\mu_3^2+b_D\mu_4^2)}{9}$ & $0$ & $\frac{4b_Dm_{\pi}^2}{3}$ & $\frac{4b_Dm_{\pi}^2}{3}$ & $\frac{(b_D-3b_F)\mu_2^2}{6}$ & $\frac{(b_D-3b_F)\mu_2^2}{6}$     \\ [2mm]             
{\bf $\eta\Sigma^0$} &                 &                 &                                      &                            &                             & $\frac{4(b_0\mu_3^2+b_Dm_{\pi}^2)}{3}$ & $\frac{4b_Fm_{\pi}^2}{\sqrt{3}}$ & $\frac{-4b_Fm_{\pi}^2}{\sqrt{3}}$ & $\frac{-(b_D+b_F)\mu_2^2}{2\sqrt{3}}$ & $\frac{(b_D+b_F)\mu_2^2}{2\sqrt{3}}$  \\ [2mm]             
{\bf $\pi^+\Sigma^-$}&                 &                 &                                      &                            &       &               & $4(b_0+b_D)m_{\pi}^2$ & $0$ & $(b_D+b_F)\mu_1^2$ & $0$ \\
{\bf $\pi^-\Sigma^+$}&                 &                 &                                      &                            &           &               &                       & $4(b_0+b_D)m_{\pi}^2$ & $0$ & $(b_D+b_F)\mu_1^2$  \\ [2mm]
{\bf $K^+\Xi^-$}     &     &     &     &       &       &        &     &     & $4(b_0+b_D)m_K^2$ & $2(b_D-b_F)m_K^2$   \\ [2mm]
{\bf $K^0\Xi^0$}     &     &     &               &       &       &               &     &     &    & $4(b_0+b_D)m_K^2$  \\ [2mm]
\hline
\multicolumn{11}{c}{} \\[15mm]
\multicolumn{11}{c}{{\rm $L_{ij}$ coefficients}} \\[2.5mm]
\hline
 & & & & & & & & & & \\[-2.5mm]
  &{\bf $K^-p$}&{\bf $\bar{K}^0n$}&{\bf $\pi^0\Lambda$}& {\bf $\pi^0\Sigma^0$}&{\bf $\eta\Lambda$}&{\bf $\eta\Sigma^0$}&{\bf $\pi^+\Sigma^-$}&{\bf $\pi^-\Sigma^+$}&{\bf $K^+\Xi^-$}&{\bf $K^0\Xi^0$}  \\
 \hline
& & & & & & & & & & \\[-2.5mm]
{\bf $K^-p$}         & $2d_2+d_3+2d_4$ & $d_1+d_2+d_3$   & $\frac{-\sqrt{3}(d_1+d_2)}{2}$ & $\frac{-d_1-d_2+2d_3}{2}$  &             $\frac{d_1-3d_2+2d_3}{2}$ & $\frac{d_1-3d_2}{2\sqrt{3}}$ & $-2d_2+d_3$ & $-d_1+d_2+d_3$ & $-4d_2+2d_3$ & $-2d_2+d_3$   \\ [2mm] 
{\bf $\bar{K}^0n$}   &                 & $2d_2+d_3+2d_4$ & $\frac{\sqrt{3}(d_1+d_2)}{2}$  & $\frac{-d_1-d_2+2d_3}{2}$  &         $\frac{d_1-3d_2+2d_3}{2}$ & $\frac{-(d_1-3d_2)}{2\sqrt{3}}$ & $-d_1+d_2+d_3$ & $-2d_2+d_3$ & $-2d_2+d_3$ & $-4d_2+2d_3$  \\ [2mm] 
{\bf $\pi^0\Lambda$} &                 &                 & $2d_4$                         &  $0$                       &          $0$ & $d_3$  & $0$ & $0$ & $\frac{\sqrt{3}(d_1-d_2)}{2}$ & $\frac{-\sqrt{3}(d_1-d_2)}{2}$   \\ [2mm] 
{\bf $\pi^0\Sigma^0$}&                 &                 &                                & $2(d_3+d_4)$               &         $d_3$ & $0$ & $-2d_2+d_3$ & $-2d_2+d_3$ & $\frac{d_1-d_2+2d_3}{2}$ & $\frac{d_1-d_2+2d_3}{2}$ \\ [2mm] 
{\bf $\eta\Lambda$}  &                 &                 &                                &                            &         $2(d_3+d_4)$ & $0$  & $d_3$ & $d_3$ & $\frac{-d_1-3d_2+2d_3}{2}$ & $\frac{-d_1-3d_2+2d_3}{2}$   \\ [2mm]              
{\bf $\eta\Sigma^0$} &                 &                 &                                &                            &         & $2d_4$ & $\frac{2d_1}{\sqrt{3}}$&$\frac{-2d_1}{\sqrt{3}}$&$\frac{-(d_1+3d_2)}{2\sqrt{3}}$ & $\frac{d_1+3d_2}{2\sqrt{3}}$  \\ [2mm]              
{\bf $\pi^+\Sigma^-$}&                 &                 &                                &                            &         &        & $2d_2+d_3+2d_4$ & $-4d_2+2d_3$ & $d_1+d_2+d_3$ & $-2d_2+d_3$   \\ [2mm] 
{\bf $\pi^-\Sigma^+$}&                 &                 &                                &                            &         &        &                 & $2d_2+d_3+2d_4$ & $-2d_2+d_3$ & $d_1+d_2+d_3$  \\ [2mm] 
{\bf $K^+\Xi^-$}     &     &     &     &       &       &        &     &     & $2d_2+d_3+2d_4$   & $-d_1+d_2+d_3$   \\ [2mm] 
{\bf $K^0\Xi^0$}     &     &     &     &       &       &       &     &     &    & $2d_2+d_3+2d_4$  \\ [2mm] 

\hline
\end{tabular}
\end{sideways}
\caption{$D_{ij}$ and $L_{ij}$ coefficients of Eq.~(\ref{V_NLO}).}
\label{tab8}
\end{table*}

%%%%%%%%%%%%%%%%%%%%%%%%%%%%%%%%%%%%%%%%%%%%%%%%

%\newpage

\end{document}